\documentclass[aps,superscriptaddress,10pt,nofootinbib,notitlepage,twocolumn,prd]{revtex4-2}
\usepackage{hyperref}
\usepackage{comment}
\usepackage{graphicx}
\usepackage[dvipsnames]{xcolor}
\usepackage{dcolumn}
\usepackage{bm}
\usepackage{amsmath}
\usepackage{amssymb}
\usepackage{color}
\usepackage{float}
\usepackage[normalem]{ulem}

\newcommand{\bx}{\boldsymbol{x}}
\newcommand{\bk}{\boldsymbol{k}}
\newcommand{\brho}{\bar\rho}
\newcommand{\tdB}{t_{\rm dB}}
\newcommand{\kp}{k_*}
\newcommand{\ks}{k_{\rm s}}
\newcommand{\dB}{{\rm dB}}
\newcommand{\SI}{{\rm SI}}
\newcommand{\QM}{{\rm dB}}
\newcommand{\GN}{{\rm N}}
\newcommand{\vg}{g} 
\newcommand{\therm}{{\rm therm}}
\newcommand{\ttherm}{t_\therm}
\newcommand{\sigshell}{\sigma_{\rm s}}
\newcommand{\cc}{{\rm c.c.}}
\newcommand{\FDM}{FDM}
\newcommand{\SIFDM}{SIFDM}
\newcommand{\rms}{{\rm RMS}}
\begin{document}


\title{Wave Interference in Self-Interacting Fuzzy Dark Matter}

\author{Christian Capanelli}
\affiliation{Department of Physics \& Trottier Space Institute,
McGill University, Montr\'eal, QC H3A 2T8, Canada.}

\author{Wayne Hu}
\affiliation{Kavli Institute for Cosmological Physics and Enrico Fermi Institute, The University of Chicago, Chicago, IL 60637, USA}

\author{Evan McDonough}
\affiliation{Department of Physics, University of Winnipeg, Winnipeg MB, R3B 2E9, Canada}

\begin{abstract}
In the Fuzzy Dark Matter (FDM) scenario, the dark matter is composed of an ultra-light scalar field with coherence length and wave interference on astrophysical scales. 
Scalar fields generically have quartic self-interactions that modify their dispersion relation and the associated evolution of density perturbations. We perform the first dedicated analysis of the role of wave interference on this evolution due to self-interactions in FDM and vice versa,
developing  a perturbative treatment applicable 
at early times and then comparing against a suite of fully nonlinear 
benchmark simulations, varying the dark matter density, interaction strength, and fiducial momentum scale.  We explicitly simulate the limit where this momentum scale is relatively high compared with the scale of the simulation volume, applicable to cases where the dark matter is initially ``warm" due to causal constraints on a post-inflationary production or in virialized halos and other ``thermalized" cases with initially cold production.  We find that in such scenarios, density perturbations are unable to grow on the expected self-interaction time scale because of interference effects, instead saturating on the much shorter de Broglie crossing time, with a dependence on the sign of the interaction. Finally, we comment on the implications of our results for astrophysical systems such as high-density ultra-faint dwarf galaxies where wave interference plays an important role.
\end{abstract}

\maketitle


\section{Introduction}
\label{sec:intro}

Among the many proposed dark matter candidates, Fuzzy Dark Matter (\FDM) 
stands out as novel
both in its theory formulation and in the avenues for detection.  Here the dark matter is so light that its de Broglie wavelength can be on  astrophysical
scales, leading to new testable predictions for dark matter (for reviews see e.g.\ \cite{Ferreira:2020fam, Hui:2021tkt}). While the original \FDM\ incarnation was  motivated by observations as a resolution to small scale problems of the $\Lambda$CDM model \cite{Hu:2000ke}, the general class of models finds further motivation from theory, as a means to probe the axiverse of string theory \cite{Svrcek:2006yi,Arvanitaki:2009fg,Cicoli:2012sz} and field theory \cite{Maleknejad:2022gyf,Alexander:2023wgk,Alexander:2024nvi}.
Although the label ``fuzzy" often is reserved for  particle masses $m\sim10^{-22} \ {\rm eV}$, we will henceforth refer to these generalizations where the de Broglie wavelength is astrophysically large simply as FDM. This generalization parallels the developments of axions themselves. Initially motivated by observations as a solution to the strong CP problem, and remaining largely successful at doing so, axions and axion-like particles have grown into a prominent dark matter candidate with broad theory motivations.

\FDM\ is distinguished from conventional particle dark matter by its novel observable imprints across cosmic epochs: in the early universe from the cosmic microwave background \cite{Hlozek:2014lca,Poulin:2018dzj,Urena-Lopez:2023ngt,Liu:2024yne} and growth of density perturbations  \cite{Dentler:2021zij,Lague:2021frh},
in the late universe from modifications to dynamical friction \cite{Lancaster:2019mde}, stellar kinematics of ultrafaint dwarfs \cite{Hayashi:2021xxu}, and by halo (sub)structure \cite{Du:2016zcv,Alexander:2019qsh,Chan:2021bja,Elgamal:2023yzt}, to present day particle physics experiments (e.g.\ \cite{Brdar:2017kbt}).  Other observable tests includes gravitational wave lensing \cite{Singh:2025uvp}, lensing of stars \cite{Eberhardt:2025tao}, modifications to the physics of 21cm cosmology \cite{Nebrin:2018vqt} and pulsar timing arrays \cite{Eberhardt:2024ocm},
along with the suite of tests of axions generally. 

Despite being wavelike, \FDM\ behaves as particle dark matter on scales much greater than their de Broglie wavelength, though even here interference phenomena play a hidden role.  Large halos still collapse and virialize but the multistream coarse grain velocity dispersion ($\sigma_v$) of collisionless particle dark matter is replaced by an isotropic distribution of wave momenta $\kp \sim  m \sigma_v$ whose wavelength $2\pi /\kp$ is much shorter than the size of the halo and whose quantum pressure behaves like velocity dispersion.  The free streaming and interference of these waves is what prevents their density fluctuation from further growing under gravity. These considerations also apply to \FDM\ produced after inflation where the horizon at production sets the characteristic wave momenta $\kp$
\cite{Amin:2022nlh,Liu:2024pjg,Ling:2024qfv}.  These interference effects are characterized by the de Broglie timescale: the wave crossing time of a de Broglie wavelength $t_{\dB}= m/\kp^2$.  Even though gravity acts over long time scales of the dynamical time $t_\GN =(G\rho)^{-1/2}$, wave free streaming and interference effects prevent gravitational collapse in the same way as velocity dispersion for particle dark matter.
On long timescales and large physical scales, these interference effects are indistinguishable from particle dark matter.

On smaller scales, wave interference is the hallmark, or {\it smoking gun} for the detection of \FDM. 
Interference is ubiquitous in \FDM\ cosmology, from dwarf galaxies to cosmic filaments  \cite{Desjacques:2017fmf,Mocz:2019pyf,Mocz:2019uyd,Dome:2022eaw,Dome:2023ska,Zimmermann:2024vng}. Interference can also lead to the generation of vortices \cite{Hui:2020hbq,Brax:2025uaw,Brax:2025vdh} which have their own set of observables \cite{Alexander:2021zhx,Alexander:2019puy}. The appearance of wave interference effects distinguishes these phenomena from  other predictions of \FDM, such as suppression of the halo mass function, which can be mimicked by warm dark matter and  particle interactions \cite{Schutz:2020jox}.

The early successes and rich range of observable phenomena paint a promising future for \FDM\ as a paradigm.  However the specific incarnations of the paradigm are  not without their challenges, and indeed its role in resolving various small-scale structure issues in the $\Lambda$CDM paradigm has come under increasing pressure from the satellite galaxies \cite{Safarzadeh:2019sre}, from the Lyman-alpha forest \cite{Rogers:2020ltq}, from strong lensing \cite{Powell:2023jns}, and other probes. Among the strongest constraints to date and one that targets its hallmark prediction is that from the dynamical heating of stellar orbits in ultrafaint dwarf galaxies that would result from wave interference in systems where $t_\GN \sim t_\dB$ \cite{Dalal:2022rmp}. These challenges have motivated generalizations and extensions of the \FDM\ model which retain the successes of \FDM\ and satisfy observational constraints. For example,  {\it self-interacting} fuzzy dark matter (SIFDM) (see e.g.~\cite{2011PhRvD..84d3531C,Chavanis:2017loo,Suarez:2017mav}) has been argued (by e.g. Ref.~\cite{Mocz:2023adf,Painter:2024rnc}) to be a \FDM\ model that retains the key successes of fuzzy dark matter while surmounting the challenges.\footnote{Multi-component fuzzy dark matter has also been proposed with this aim, see e.g., \cite{Eby:2020eas}.}

In parallel with this, a growing body of work has studied the impact of self-interactions in systems involving light scalar fields, uncovering significant physical impacts and qualitative changes to the physics. For example, in the context of black hole superradiance \cite{Baryakhtar:2020gao}, self-interactions of the size typical for axion dark matter are sufficient to shut off the superradiant instability, with the occupation number of axions saturated to a quasi-equilibrium level.  

Similarly, a small self interaction can affect the global structure of a dark matter halo since it acts over a long time scale, similar to gravitational interactions (e.g. \cite{Amin:2019ums, Mocz:2023adf}). Relatedly,
for the QCD axion, despite possessing an extremely small self-interaction given the dark matter abundance, it has recently been claimed by Ref.~\cite{Gorghetto:2024vnp} that the collapse of small scale objects is qualitatively changed due to evolution over the interaction timescale by changing both the momentum distribution and density perturbations of the axions.

On the other hand, just as the interplay between the de Broglie timescale $t_\dB$ and the gravitational dynamical timescale $t_\GN$ involves wave interference phenomena to establish the free streaming stability scale, interference effects can play a role in the impact of self-interactions as well, even when their relevant time scale $t_\vg$ is hierarchically separated from either.
 We note that wave interference effects are neglected by simulations that treat \SIFDM\ as a fluid (such as \cite{Dawoodbhoy:2021beb}).
 The exact correspondence to a fluid  breaks down at singular points in the density field, whereas the wave function remains well defined. Thus the impact of  
 wave interference on the dynamics of self-interaction remains an open problem.

In this work we perform the first dedicated analysis of wave interference in self-interacting fuzzy dark matter (\SIFDM). Using a combination of analytical tools and numerical simulations, we identify qualitatively distinct dynamics that depend on the initial momentum spectrum of the waves. We show that long wavelength density fluctuations generated by interference of short wavelength (high momenta) field fluctuations do not grow over the naive self interaction timescale, because the growth requires coherence of the interacting waves that is only maintained on a much shorter timescale. In contrast, for long wavelength (low momentum) field fluctuations, self-interactions do change the density spectrum on the usual time scale.   This behavior occurs in the QCD axion collapse recently studied in Ref.~\cite{Gorghetto:2024vnp}.

To show this, we develop a perturbative approach to wave interactions, which is general and flexible enough to describe both regimes. We make predictions for the field and density power spectra which are confirmed by numerical simulations.  We use this to clarify the conditions for which the two types of behaviors exist.

The outline of this paper is as follows: in Sec.~\ref{sec:SIFDM} we provide a general overview of \SIFDM\ including the characteristic timescales of the model. In Sec.~\ref{sec:PT} we develop the perturbation theory for interacting waves, and in Sec.~\ref{sec:sims} we perform a suite of numerical simulations. We find excellent agreement between numerical simulations and perturbations theory within the domain of validity of the latter. We conclude in Sec.~\ref{sec:discussion} with a discussion of implications and applications of this work.

\section{Self-Interacting FDM}
\label{sec:SIFDM}

Fuzzy dark matter originates from relativistic scalar field theory,  described by the action
\begin{equation}\label{eq:action}
 S=\int \mathrm{d}^4 x \sqrt{-g}\left[\frac{1}{2}\left(\partial_\mu \phi\right)^2-\frac{m^2}{2}\phi^2-\frac{\lambda}{4!} \phi^4\right]
\end{equation}
in units where $\hbar=c=1$.
The self-interaction term $\lambda\phi^4$  is a dimension-4 operator that is allowed and therefore expected in effective field theory, and is present in many examples of scalars in particle physics, both discovered (the Standard Model Higgs) and yet to be discovered (e.g., axion-like particles), and in condensed matter physics, where the self-interaction plays a crucial role in superfluidity.

In the realm of ultralight dark matter, a natural candidate for $\phi$ is an axion-like particle. The self-interaction $\lambda$ is expected to be small and attractive in conventional axion-like models where $V(\phi ) = m^2 f ^2 \left[ 1- \cos(\phi/f)\right]$ and hence $\lambda= - m^2/f^2$. For example,  QCD
axion dark matter with  $m\sim 10^{-6}$ eV and decay
constant $f_a \sim 10^{12}$ GeV yields $\lambda\sim 10^{-54}$. However, $\lambda$ can differ in sign and size in  
concrete realizations of ultralight or fuzzy dark matter.
A simple explicit working model of an axion-like particle with independent and tuneable $m$, $f$, and $\lambda$, is given by the potential
\begin{equation}
V = m^2 f^2 \left[ 1- \cos(\phi/f) \right] +  \frac{1}{6}\lambda_0 f^4 \left[ 1 - \cos(\phi/f) \right]^2,
\end{equation}
motivated by the generalized axion potential proposed in \cite{Kamionkowski:2014zda} and later applied to the Hubble tension problem in the context of Early Dark Energy \cite{Poulin:2018cxd}. This
leads to an effective self-interaction $\lambda_{\rm eff}= \lambda_0 - m^2/f^2$. A UV complete description of this model is provided by a minimal modification to the string theory models for Early Dark Energy presented in Refs.~\cite{McDonough:2022pku,Cicoli:2023qri}. In these models $\lambda_0$ can be positive or negative, and both $m$ and $\lambda$ are generated by non-perturbative effects and are radiatively stable.  Other concrete examples of self-interacting ultralight scalars can be found in e.g.\ \cite{Fan:2016rda}. 

With all this in mind, in this work we treat the self-interaction as a free parameter to constrained by observation  \cite{Li:2013nal,Cembranos:2018ulm,Chavanis:2020rdo, Hartman:2021upg,Chakrabarti:2022owq,Dave:2023egr, Aboubrahim:2024spa}
and remain agnostic about the possible UV origins of this interaction. 

To be a viable dark matter candidate, FDM
must be highly non-relativistic at the present.   The dark matter can be still be ultralight  if it is produced nonthermally.   In this regime, particle modes are highly occupied and the scalar field is well described by the  Schr\"odinger equation for the wavefunction $\psi$ \cite{Salehian:2020bon,Salehian:2021khb,Zhang:2024bjo}:
\begin{equation}
     \phi = \frac{1}{\sqrt{2m}}\left(\psi e^{-imt}+\cc     \right)
\end{equation}
as long as $|\ddot{\psi}|\ll m |\dot{\psi}|$.  Including gravitational and self interactions through the Newtonian potential $V_\GN$ and self-interaction potential $V_\SI$, we obtain
\begin{eqnarray}\label{eq:schro-poisson}
    i\dot{\psi}&&= -\frac{1}{2m}\nabla^2\psi+V_\GN\psi+V_\SI \psi, \\
    \nabla^2 V_\GN &&= 4\pi mG\rho, \quad 
    V_\SI = \frac{g \rho}{2m} ,\nonumber
\end{eqnarray}
where $\rho$ is the mass density,
\begin{equation}
\label{eq:rho}
    \rho({\bx},t) \equiv m |\psi({\bx},t)|^2,
\end{equation}
and $g$ is a coupling constant of mass dimension $-2$, related to parameters in the underlying relativistic theory by $g= \lambda/(4m^2)$. When $g=0$, the model is simply FDM, when $g>0$ ($g<0$) it is repulsive (attractive) \SIFDM.  
We assume throughout that the relevant timescales are much shorter than the cosmological expansion time.

From  Eq.~\eqref{eq:schro-poisson} one can immediately identify the characteristic timescales for evolution of the wavefunction.  For waves with a characteristic momentum scale  $\kp$, one may identify a timescale for free propagation   
\begin{equation}
t_\dB = \frac{m}{\kp^2},
\label{eq:tdB}
\end{equation}
and  for interaction
\begin{equation}
\label{eq:tv}
t_\vg = \frac{m}{|g| \brho},
\end{equation}
where 
the overbar denotes a spatial average.
On the other hand, in Eq.~(\ref{eq:schro-poisson}) these timescales explicitly relate to the evolution of the phase of $\psi$,
or equivalently the energy-momentum and dispersion relations of the particles or waves respectively \cite{Semikoz:1995rd},
as opposed to changes in its momentum distribution and or amplitude which determines  the spatial density $\rho$.  

These other timescales can be better identified once we perform a change of variables from a single complex field into two real fields 
\begin{equation}
\psi(\boldsymbol{x},t)=
\sqrt{\frac{\rho(\boldsymbol{x},t)}{m}}e^{i\theta(\boldsymbol{x},t)}
\end{equation}
and recast   the Schr\"odinger-Poisson system \eqref{eq:schro-poisson} as 
a set of hydrodynamical equations:
\begin{align}\label{eq:euler-madelung}
    \dot{\rho}+\nabla \left( \rho \boldsymbol{v}\right)&=0,\\
    \dot{\boldsymbol{v}}+\left(\boldsymbol{v}\cdot \nabla\right)\boldsymbol{v}&=-\frac{\nabla}{m}\left(V_{\QM}+V_{\GN}+V_{\SI} \right), \nonumber
\end{align}
where $\boldsymbol{v}\equiv -\nabla \theta / m$ is the bulk fluid velocity, and
\begin{align}
&V_{\dB} = -\frac{1}{2m}\frac{\nabla^2\sqrt{\rho}}{\sqrt{\rho}} 
\end{align}
is the effective quantum pressure. 
For each type of potential $V$ when considered alone, we can define a timescale associated the change in the density
\begin{equation}
\frac{\ddot \rho}{\rho} = {\cal O}( t_V^{-2})
\end{equation}
such that
\begin{equation}
t_V  \equiv \frac{1}{\sqrt{|\nabla^2 V|/m}}.
\end{equation}
Notice that for the Newtonian potential,
\begin{equation}
t_\GN = \frac{1}{\sqrt{ G \rho}},
\end{equation}
which is the usual dynamical time, and if we 
 further associate a dominant Fourier wavenumber $\kp$ for the spatial derivatives of $V_\QM$ we have 
\begin{equation}
t_{\dB} = \frac{m}{\kp^2},
\end{equation}
which is the de Broglie time. 
Then for the self interaction
\begin{equation}\label{eq:tSI}
t_{\SI} =\frac{1}{\sqrt{\nabla^2 V_{\SI}/m}} = \sqrt{t_\dB t_\vg}.
\end{equation}
For convenience, we can define the dimensionless quantity
\begin{equation}
\epsilon_g  = \frac{g\rho}{\kp^2},
\end{equation}
such that $t_{\SI}/t_\vg =\sqrt{|\epsilon_g|}$.  When $|\epsilon_g| \ll 1$, the Schr\"odinger equation is in the so-called kinetic regime and self-interactions provide a slow correction
to the time evolution and a small correction to the energy-momentum relation \cite{Semikoz:1995rd}.

Finally, as we shall discuss in the next section, the timescale $t_\therm = t_\vg/|\epsilon_g|$ is associated with the non-perturbative effects of interactions. 

For many physical situations, these timescales will be hierarchically separated, leading to the dominance of free wave, self-interaction, or gravitational effects.  On the other hand, these timescales only appear when considering a single  dominant effect in isolation and its impact on a single characteristic momentum scale $\kp$.   Over long timescales, subdominant effects can change the momentum and density distribution. 
In fact, for the case of slow gravitational collapse under wave evolution in FDM, the hydrodynamic picture breaks down where wave interference produces density nodes, since  $V_\QM$ becomes undefined for $\rho(\bx,t)=0$. In the particle picture, this indicates shell crossing has occurred and the subsequent multistream behavior can lead to virial equilibrium from the coarse grain velocity dispersion, rather than continued collapse on the $t_\GN$ timescale.  This particle picture is replicated by wave dynamics in FDM simulations on scales much larger than  $\kp^{-1}$ but is instead interpreted as the effect of interference \cite{Gough:2022pof,Gough:2024gim}.

The combination of wave interference and self interaction is less well studied. We develop a perturbative framework to consider these cases next.

\section{Wave Perturbation Theory}
\label{sec:PT}

In this section, we develop a  methodology for perturbatively treating wave-wave interactions and their impact on the momentum distribution and density power spectrum of those waves.
Here we consider the impact of the self-interactions and neglect gravitational interactions, i.e.\ we assume $t_\GN \gg {\rm max}(t_\dB,t_\vg)$ and evolution across $t\ll t_\GN$,  but the formalism can be readily  extended to the converse.

In Sec.\ \ref{sec:greenwave}, we set up the perturbative expansion and in Sec.\ \ref{sec:fieldpower} and Sec.\ \ref{sec:densitypower} we evaluate the leading order contributions to the field and density power spectra respectively.

\subsection{Wave Interactions}
\label{sec:greenwave}

For the background around which we perturb, we take the case where the volume average density $\brho$ is dominated by field momenta whose wavelengths are much shorter than the averaging scale.  This occurs, for example, within dark matter halos due to their virial velocity, or for an FDM production mechanism in the early universe after inflation due to the small causal horizon.
We take an initial distribution  $\psi_i(\bk) = \psi(\bk,t=0)$ determined by the power spectrum
\begin{equation}
\langle \psi_i^*({\bk}) \psi_i({\bk}')\rangle = (2\pi)^n \delta({\bk}-{\bk}') P_{i}(k),
\label{eq:Pi}
\end{equation}
where we have defined the momenta in $n$ spatial dimensions with the Fourier transform for the field
\begin{eqnarray}
\psi(\bk,t) &=& \int d^n x e^{-i \bk \cdot\bx  }\psi(\bx,t) 
\end{eqnarray}
and for the density modes
\begin{eqnarray}
\rho(\bk,t) &=&  m \int  \frac{d^n k'}{(2\pi)^n} \psi^*(\bk'-\bk,t) \psi(\bk',t).
\label{eq:densitymode}
\end{eqnarray}
Furthermore, the initial density power spectrum is given by the initial field spectrum as
\begin{eqnarray}
\langle \rho^*(\bk,0)\rho(\bk',0)\rangle
&=& (2\pi)^n \delta({\bk}-{\bk}') P_{\rho}(k,0), \\
P_{\rho}(k,0)  &=&
m^2 \int \frac{d^n k_a}{(2\pi)^n} 
P_i({\bk}_a-{\bk})P_i({\bk}_a).\nonumber
\end{eqnarray}
Note that the spatial average over a volume $V$ is $\brho = \rho(\bk=0,t) /V$.
Spatial averages, denoted by the overbar, in general differ from ensemble averages, denoted by brackets.  In this section we keep these distinct, but the distinction will not be important for the timescale analysis here or the simulations in the next section. 

Although we always consider $3$ spatial dimensions in the action of Eq.~\eqref{eq:action}, we leave $n$ arbitrary here so as to allow the initial momentum spectrum to be populated only along a subset $n\le 3$ so as to ease the computational burden of simulations below.   For example if the initial field has no $z$ dependence then the momenta are restricted to $k_x,k_y$ for all time via the Schr\"odinger dynamics.

Since the total number of particles and $\bar \rho$ is conserved,
we can immediately solve the Schr\"odinger equation for a field interacting with the mean density, which we consider as the ``unperturbed solution"
\begin{equation}
\psi_0(\bk,t) = \psi_i(\bk) e^{-i\frac{k^2 + g\brho}{2m} t}.
\label{eq:psi0}
\end{equation}
In this case, the initial field just picks up a time-dependent phase on a timescale of  $t_\dB$ and $t_\vg$ for free propagation and interaction respectively for any fiducial $\kp=k$ (see Eq.~\ref{eq:tdB}, \ref{eq:tv}).  On the other hand, the amplitude does not evolve, nor do the statistical properties of the density field that $\psi_0$ generates.

Therefore the non-trivial effect of the coupling $g$ is through the interactions with density {\it fluctuations} from the local source term 
\begin{equation}
S(\bx,t)=\frac{g}{2m} (\rho(\bx ,t)-\brho) \psi(\bx,t).
\end{equation}
Using  Eq.~(\ref{eq:schro-poisson}),  these interactions perturb the field away from Eq.~(\ref{eq:psi0}), $\psi = \psi_0+\delta\psi$ through
\begin{eqnarray}
    i\dot{\delta\psi}&&= -\frac{1}{2m}\nabla^2\delta\psi+S(\bx ,t).
    \label{eq:psiiteration}
\end{eqnarray}
We can then solve for the perturbation iteratively 
\begin{equation}
\psi = \psi_0 + \delta\psi=\psi_0+\psi_1 + \ldots
\end{equation}
by using the $n$th order solution for $\psi$ to evaluate $S$ and find the ($n+1$)th order correction to $\psi$ using the Green function approach for considering  $S_n$ as an external source.  In particular the first order correction is 
\begin{eqnarray}
\psi_1(\bk,t) &=&  -i\frac{g}{2}  e^{-i\frac{k^2 + g\brho}{2m} t}\left[ \int_0^t dt' e^{i \frac{k^2}{2m} t' }
\right. \nonumber\\
&& \times \left.
 \int\frac{d^n k_a}{(2\pi)^n}
 \int \frac{d^n k_b}{(2\pi)^n}
e^{i \frac{k_c^2 - k_a^2 - k_b^2}{2m} t'} \right. \\
&& \times \left.  \psi_i(\bk_a) \psi_i(\bk_b) \psi_i^*(\bk_c)  - \frac{\brho}{m} t \psi_i(\bk) \right], \nonumber
\end{eqnarray}
where  
$\bk_c=\bk_a+\bk_b -\bk.$ 
This series is heuristically an early time expansion in orders of
\begin{equation}
\frac{g}{2m} (\rho(\bx ,t)-\bar\rho) t \ll 1
\end{equation}
and as we shall see, the accuracy of first order perturbation theory at some time $t$ will depend on the momentum spectrum $P_i(k)$ that determines $\rho(\bx, t)$.  There is no explicit requirement that $g\brho t/m= t/t_\vg \ll 1$, nor  $t_\vg/t_\dB=|\epsilon_g|\ll 1$, since  the background effect is  non-perturbatively included in $\psi_0$ as a simple phase.   For shorthand however, we will count orders by powers in $g$ that appear in the scalings of amplitudes, e.g.\ $\psi_1 = {\cal O}(g)$.

\subsection{Field Power Spectrum}
\label{sec:fieldpower}

We  next consider the leading order change in the field power spectrum and estimate the timescale over which these changes remain perturbative.
If we denote the contributions at ${\cal O}(g^{a+b})$ from 
\begin{equation}
\langle \psi_a^*({\bk},t) \psi_{b}({\bk}')\rangle = (2\pi)^n \delta({\bk}-{\bk}') P_{ab}(k,t),
\end{equation}
we have no contribution to ${\cal O}(g)$, since
 $P_{10} = P_{01}^*$ by construction and both are pure imaginary.
The leading order term is ${\cal O}(g^2)$
\begin{equation}
\delta P= P - P_i = P_{11} + P_{20} + P_{02}+ \ldots
\end{equation}
Moreover, if the initial state is unpopulated for a given $k$ mode $\psi_i(k)=0$ then $\psi_0(k,t)=0$, $P_{20}=P_{02}=0$, and  $\delta P = P_{11}$  to leading order.  If the connected correlators higher than the two point in Eq.~(\ref{eq:Pi}) vanish initially then we obtain
\begin{eqnarray}
\label{eq:P11}
P_{11}(k,t) &=& \frac{g^2}{2}  \int\frac{d^n k_a}{(2\pi)^n}
 \int \frac{d^n k_b}{(2\pi)^n} P_i(k_a) P_i(k_b) P_i(k_c) \nonumber\\
 &&\times\left[\frac{4 m}{\Delta k^2} \sin\left( \frac{ \Delta k^2}{4m} t \right)\right]^2,
\end{eqnarray}
with
\begin{equation}
\Delta k^2 \equiv k^2 + k_c^2 - k_a^2 - k_b^2,
\label{eq:Deltak2}
\end{equation}
where recall
\begin{equation}
\bk_c=\bk_a+\bk_b -\bk.
\label{eq:kc-field}
\end{equation}  
This relatively simple expression contains a wealth of information, regarding both the time and $k$ dependence of the evolution.   Notice that $\Delta k^2$ represents the phase coherence between the 3 initial modes $k_{a,b,c}$ that generate $\psi_1(k)$ and the phase of this mode itself (see Fig.~\ref{fig:modes} below).   The interaction causes coherent growth of $P_{11}$ when $\Delta k^2 t/4 m\ll 1$.    Once this coherence is lost, $P_{11}$ stops growing and instead oscillates around some average value.

We can get further insight into this generic behavior in the case where the mode $k \ll \kp$, the typical momentum in the initial spectrum.  Here
\begin{equation}\label{eq:delta-k-sq}
\lim_{k\ll \kp} \Delta k^2 =k_c^2-k_a^2-k_b^2 \sim - \kp^2,
\end{equation}
and the decoherence time is related to $t_\dB$, the timescale associated with the free-field phase evolution itself.  At early times
\begin{equation}
\label{eq:P11integral}
\begin{split}
\lim_{t\ll t_\dB} P_{11}(k,t) \approx & \frac{g^2 t^2}{2}  \int\frac{d^n k_a}{(2\pi)^n}
 \int \frac{d^n k_b}{(2\pi)^n} \\
 & \times P_i(k_a) P_i(k_b) P_i(k_c)
 \\ \approx & \frac{g^2 t^2}{2m^2} \int \frac{d^n k_a}{(2\pi)^n} P_i(k_a) P_\rho(k_a,0).
\end{split}
\end{equation}
Notice that we can perform the angular integrals immediately, returning the area of the $n-1$ sphere, and
for any given $P_i(k)$ we can evaluate the remaining single integral over $k_a$ easily. In Sec.~\ref{sec:sims} we give an explicit example of a thin shell around $\kp$.   

On the other hand by dimensional analysis, we can extract the generic behavior by noting that $\kp^{n} P_i  \sim \brho/m$ and $\kp^{n} P_\rho \sim \brho^2$ to find
\begin{equation}
\lim_{t\ll t_\dB}\kp^n P_{11}(k,t) \sim \frac{\brho}{m}    \frac{g^2
\brho^2  }{m^2}t^2 =
 \frac{\brho}{m}\left( \frac{t}{t_\vg} \right)^2,
\end{equation}
such that the momentum spectrum grows on the $t_\vg$ timescale relative to $P_i$ until $t \sim t_\dB$, after which it saturates.   We can characterize the full time evolution  as
 \begin{equation}
\lim_{k\ll \kp} \kp^n P_{11}(k,t)\sim  \frac{\brho}{m}  \epsilon_g^2 \sin^2 (t/ 4t_{\dB}).
 \end{equation}
 Here we have dropped model dependent factors and order unity coefficients including factors of $\pi$ (cf.\ Eq. \ref{eq:field-power-theory} for their restoration in the thin shell case).  For $t_\dB \ll t_\vg$, i.e.\ $|\epsilon_g|\ll 1$, wave interference effects strongly suppress the growth and hence the change of the momentum spectrum from its initial conditions $P_i(k)$.   If $t_\dB> t_\vg$, i.e.\ $|\epsilon_g|>1$ then the growth will change the momentum occupation spectrum by $O(1)$, $P_{11}(k) \sim P_i(\kp)$ before wave incoherence can suppress further growth.   In this case perturbation theory will eventually break down and we expect substantial evolution of the spectrum on the $t_\vg$ timescale.  Note that this generically occurs when the initial momentum distribution is peaked at sufficiently small $\kp$ such that wave incoherence takes longer than the interaction time $t_\vg$ to develop.

For $k\gg \kp$ a similar analysis applies with the limitation that since 
$\bk = \bk_a +\bk_b -\bk_c$
and $k_{a,b,c} \sim \kp$,  first order perturbation theory would limit the generated mode $\psi_1$ to  $k \lesssim 3\kp$.  In this case, we still have
\begin{equation}
\Delta k^2 \sim {\cal O}(\kp^2)
\end{equation}
and the same scalings as for $k\ll \kp$ apply.  For $|\epsilon_g|\ll 1$, we again expect saturation of growth of $P_{11}$ once $t>t_\dB$ and given $k \lesssim 3\kp$ the modes with $k\gg \kp$ would remain almost unpopulated.   For  $|\epsilon_g|>1$ we would expect non perturbative evolution of the spectrum $P_i(k)$ after $t>t_\vg$ and potentially modes with $k\gg \kp$ populate through a cascade of interactions  \cite{Gorghetto:2024vnp}.

Modes where $k\sim \kp$ are  special in that they are at the characteristic momentum of  the initial spectrum and are hence highly populated.     In this case, the term
\begin{equation}
 \langle \psi_0^*(\bk,t) \psi_2(\bk',t)\rangle + \cc
 \end{equation}
is also part of the leading order expression for the evolution of $P(k\sim \kp,t)$ and these modes can continue to evolve.
  While the second order integrals are more involved for $\psi_2$, we can estimate their impact on the evolution using Eq.~(\ref{eq:psiiteration}) for $\dot\psi_2$ for the RMS fluctuation in position space.  Given
\begin{equation}
S_1 = {\cal O}(\frac{g\brho}{m})\psi_1
\end{equation}
and taking $\psi_1 \sim \epsilon_g \psi_0$ after saturation for $|\epsilon_g|\ll 1$, we can estimate 
\begin{equation}
\frac{\psi_0 ^*\dot\psi_2 + {\cc}}{|\psi_0|^2} \sim \frac{g \brho}{m} \epsilon_g  \sim \frac{ |\epsilon_g| }{t_\vg} \equiv \frac{1}{t_\therm},
\label{eq:thermtwiddle}
\end{equation}
where we have again dropped all proportionality coefficients.  This 
indicates that the initially occupied modes change by ${\cal O}(1)$ on the timescale $t_\therm$ via these higher order corrections.  This timescale has been previously identified as the ``thermalization" timescale in the kinetic regime by casting the evolution of $P(k)$ as a Boltzmann equation for the occupancy of the momentum mode $k$.   Thermalization here refers to the correspondence of $S$ to the collision term of the Boltzmann equation  where the rate of energy-momentum exchange redistributes the momentum spectrum on a timescale $t_\therm$ to establish kinetic equilibrium
\cite{Semikoz:1995rd,zakharov_kolmogorov_1992,Sikivie:2009qn,Guth:2014hsa,Kirkpatrick:2020fwd, Jain:2023ojg, Jain:2023tsr}.

\subsection{Density Power Spectrum}
\label{sec:densitypower}

\begin{figure}[t]
    \centering
\includegraphics[width=0.9\linewidth]{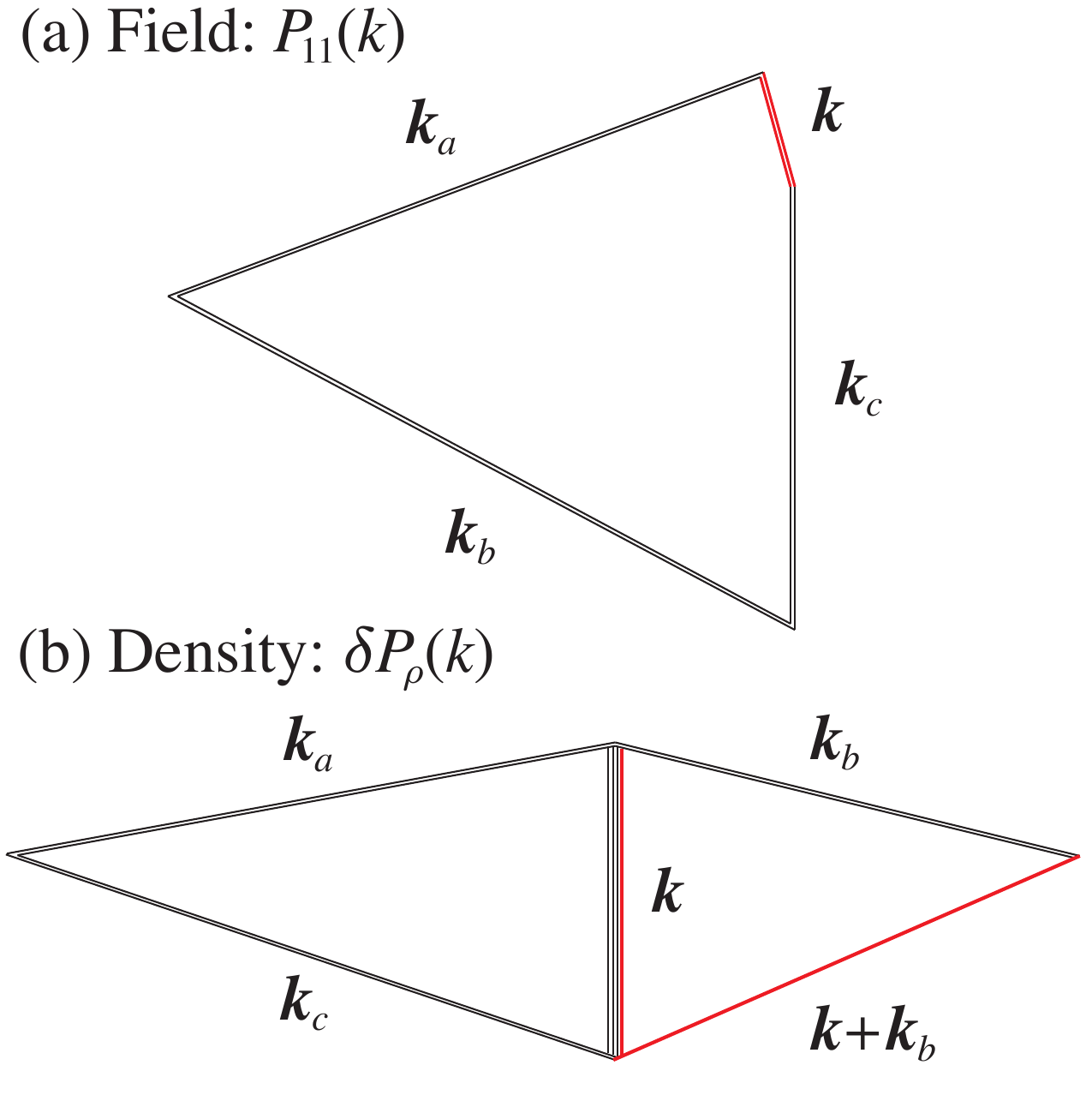}
    \caption{Construction of the perturbed power spectra for  (a)  field: $P_{11}(k)$ and  (b) density: $\delta P_{\rho}(k)$ out of $\psi({\bk'})$ field modes.  Lines represent the field modes participating in the construction with black as unperturbed modes $\psi_0$ and red as perturbed modes $\psi_1$. The number of red lines on $k$ corresponds to the order in $g$ and the total number denotes the 2 and 4pt intrinsic nature of the power spectra.  The field modes on the perimeter of the quadrilateral contribute to the phase decoherence, $\Delta k^2=k^2+k_c^2-k_a^2-k_b^2$ and $\Delta q^2=(\bk+\bk_b)^2+k_c^2-k_a^2-k_b^2$ respectively.}
    \label{fig:modes}
\end{figure}

Next let us consider the leading order evolution of the density power spectrum.
For the density spectrum, the first order contribution
 $\rho_1 = \rho-\rho_0  + {\cal O}(g^2) +\ldots$
is given by the first order field perturbation as
\begin{eqnarray}
\rho_1(\bk,t) &=& m  \int\frac{d^n k'}{(2\pi)^n} \left[ \psi_1^*(\bk',t)\psi_0(\bk'+\bk,t) \right. \nonumber\\
&& \left. + \psi_0^*(\bk'-\bk,t) \psi_1(\bk',t) \right].
\end{eqnarray}
Unlike the field power spectrum, this first order contribution produces a nonzero change in the power spectrum itself
\begin{equation}
\delta P_{\rho}(k,t) \equiv P_{\rho}(k,t)-  P_{\rho}(k,0) = {\cal O}(g),
\end{equation}
with this leading order contribution given by
\begin{equation}
 \langle \rho_0^*(\bk',t) \rho_1(\bk,t) \rangle +\cc
 = 
 (2\pi)^n \delta(\bk-\bk')\delta P_\rho(\bk,t),
 \end{equation}
where
\begin{eqnarray}\label{eq:delta-P_rho}
\delta P_\rho(\bk,t) &=& -4g m^2  \int\frac{d^n k_a}{(2\pi)^n}
 \int \frac{d^n k_b}{(2\pi)^n} P_i(k_a) P_i(k_b) P_i(k_c)\nonumber\\
 &&\times \frac{4 m}{\Delta q^2} \sin^2(\frac{\Delta q^2}{4 m} t).
\end{eqnarray}
Unlike for $P_{11}$, here  $\bk_c=\bk+\bk_a$, and
\begin{equation}
\Delta q^2 = (\bk_b+\bk)^2 + k_c^2 - k_a^2 - k_b^2 = 2(\bk_a\cdot \bk + \bk_b\cdot\bk + k^2).
\end{equation}
While the density spectrum perturbation $\delta P_\rho$ shares similar properties with the field spectrum perturbation $P_{11}$, there are notable differences.   First the density spectrum evolution is linear in $g$ and hence depends on its sign.   This is because the phase of $\psi_1$ can constructively combine with those of $\psi_0$ unlike for $P_{01}+P_{10}$.    In particular, the phase coherence term $\Delta q^2$ involves the 4 field modes that are involved in the construction of $\rho_0$ and $\rho_1$, one of which is the perturbed field mode $\psi_1(\bk_b+\bk)$ and the other three are background modes $k_{a,b,c}$ (see Fig.~\ref{fig:modes}).     We shall see that this difference means  the decoherence time depends on $k$ and  does not always occur on the $t_\dB$ timescale .   Finally, the first order expression contains all leading order contributions and unlike $P_{11}$ applies to density wavenumbers where the field modes are initially unoccupied or occupied, i.e.\ $k\sim \kp$.

Let us first consider the coherent limit where $t \ll m/\Delta q^2$
\begin{eqnarray}
\label{eq:deltaPrho}
\lim_{t \ll m/\Delta q^2} \delta P_\rho(\bk,t) &=&   -g m  t^2\int\frac{d^n k_a}{(2\pi)^n}
 \int \frac{d^n k_b}{(2\pi)^n}  \nonumber\\
&&\times      P_i(k_a) P_i(k_b) P_i(k_c) \Delta q^2\nonumber\\
&=&-g\brho   t^2\int\frac{d^n k_a}{(2\pi)^n} P_i(k_a)P_i(\bk_a+\bk) \nonumber\\
&&\times (2\bk_a\cdot \bk +k^2).
\end{eqnarray}
If we additionally take $k\ll \kp$ then we can Taylor expand $P_i(\bk_a+\bk)$ and integrate over angles analytically and over $k$ by parts assuming that the $P_i$ spectrum vanishes at $k=0,\infty$
\begin{eqnarray}
\label{eq:deltaPrholowk}
\lim_{t \ll m/\Delta q^2,k\ll \kp} \delta P_\rho(\bk,t)&=&
-\frac{g \brho k^2t^2}{m^2} P_{\rho}(k,0) \nonumber\\
&=&
-\left(\frac{k}{\kp} \frac{t}{t_\SI} \right)^2 P_{\rho}(k,0) .
\end{eqnarray}
Notice that the timescale associated with the growth of $P_\rho(k)$  is  $t = t_\SI (\kp/k)$ where $t_\SI^2=t_\vg t_\dB$ was associated with the self-interaction growth of the $\kp$ mode in Eq.~\eqref{eq:tSI}.  This rescaling reflects the same  dynamics as the fluid approximation but for the density $k$-mode in question instead of $\kp$. 

For $k \sim \kp$ the exact result will depend on the initial spectrum $P_i(k)$ but again by dimensional analysis we can infer
\begin{equation}
\lim_{t \ll m/\Delta q^2 ,k\sim \kp} \delta P_\rho(\bk,t) \sim
-\frac{t^2}{t_\SI^2} P_{\rho}(k,0)
\end{equation}
and identify the timescale as $t_\SI$ itself (cf.~Eq.~\ref{eq:deltaPrhoshell}) for the thin shell case).  Furthermore the contributing modes $k_a \sim k_b \sim k_c\sim \kp$ imply that the field perturbation mode $\psi_1(\bk_b+\bk)$ that is involved in the evolution of $P_\rho(k\sim \kp)$ has substantial contributions from field modes with $\kp \lesssim |\bk_b + \bk|  \lesssim 2 \kp$
(see also Fig.~\ref{fig:filter}).

Now let us consider the decoherence of the wavemodes and its impact on growth.   
The quadratic growth stops when $t \sim m/\Delta q^2$ and so this stopping time also depends on the configuration of the modes $\bk,\bk_a,\bk_b$ entering into $\Delta q^2$.   For $k \sim \kp$ we can see that this timescale is again $\tdB$ and
  \begin{equation}\label{eq:Prho-peak-scaling}
\lim_{t \gg \tdB} \delta P_\rho(\sim \kp,t) \sim  \frac{g\bar \rho}{\kp^2} P_\rho(\kp,0)  
= \epsilon_g  P_\rho(\kp,0) 
\end{equation}
so  if $|\epsilon_g| \ll 1$, the fractional changes do not reach ${\cal O}(1)$ on the naive timescale $t_{\rm SI}$.   The exception again is for the particular configurations where the decoherence $\Delta q^2 \rightarrow 0$ which continue to grow (see Fig \ref{fig:density-power-evolution}, right panel for a thin shell example).
 For $k \ll \kp$, the growth is slower since the timescale scales as $t \sim t_\SI (\kp/k)$ but also saturates  later since $\Delta q^2 \sim k \kp$. The combination implies that for $t\gg \tdB (\kp/k)$ we expect that eventually fractional contributions reach the same  $\epsilon_g$ suppressed saturation level unless higher order effects dominate first.  Again the consequence of wave interference is to stop the growth due to self-interactions and limit its net amount to a fraction $\epsilon_g$ when $|\epsilon_g|\ll 1$.    
We graphically summarize the similarities and differences in the construction of the field and density power spectra perturbations in Fig.~\ref{fig:modes}.

Finally notice that these scaling arguments apply independently of whether the initial field varies along all $n=3$ or $n=2$ spatial dimensions.   This independence motivates the choice of $n=2$ for the comparison with numerical simulations in the next section.

\section{Simulations}
\label{sec:sims}

We now present the results of a series of numerical simulations of \SIFDM\ in the wave interference dominated ($|\epsilon_g| \ll 1$) regime. In Sec.~\ref{sec:numericalimplemetation} we discuss the details of the simulations including initial conditions and the numerical implementation. In Sec.~\ref{sec:timedomain} we separate the time-evolution into the characteristic timescales and find excellent agreement with the perturbation theory from Sec.~\ref{sec:PT} within its domain of validity. In Sec.~\ref{sec:results} we present our results for the $k$-space field and density power spectra at these characteristic timescales.


\begin{figure*}[t]
    \centering
    \includegraphics[width=0.9\linewidth]{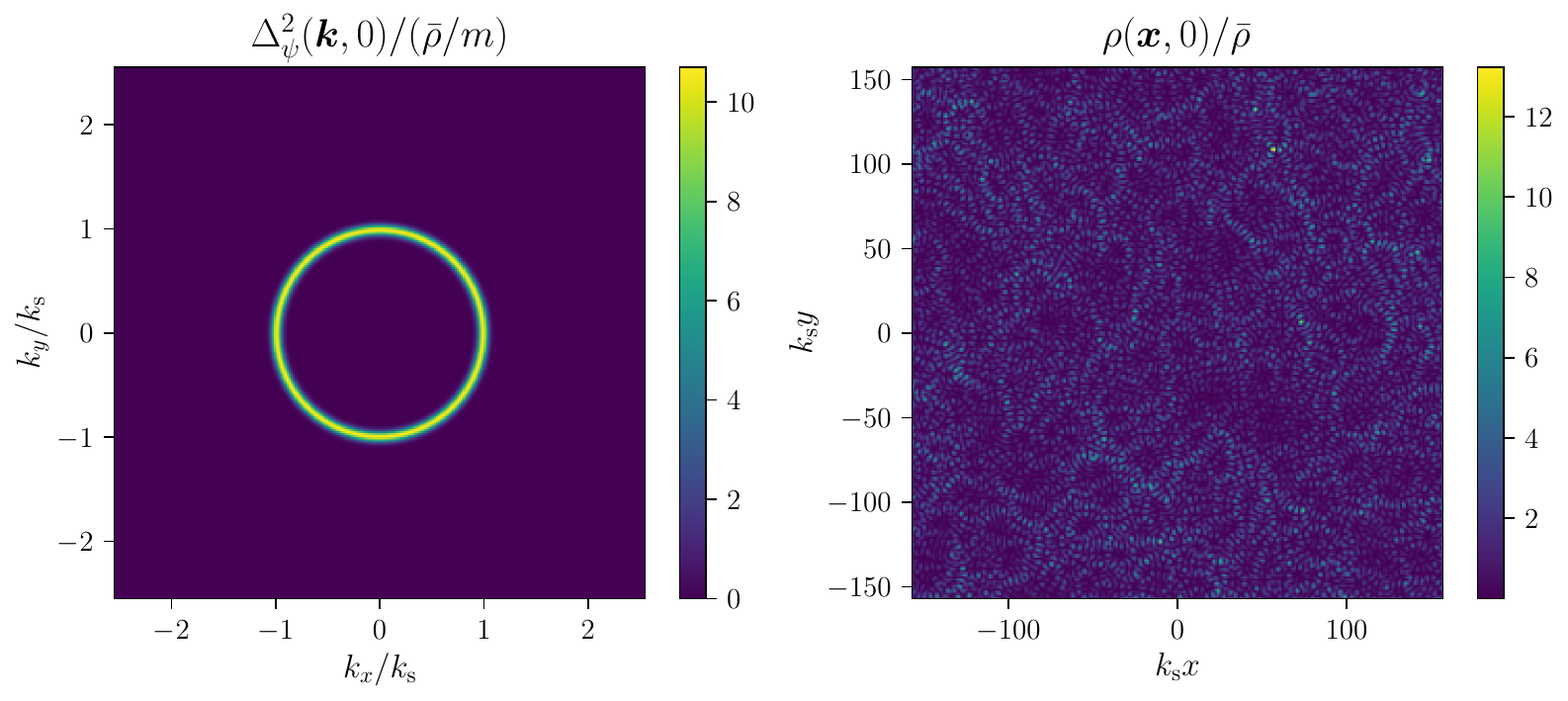}
    \caption{Example initial field power (left) and corresponding real space density (right) where the shell width is $\sigshell/\ks=0.05$.}
    \label{fig:initial-condition}
\end{figure*}

\begin{figure*}[t]
    \centering
    \includegraphics[width=0.9\linewidth]{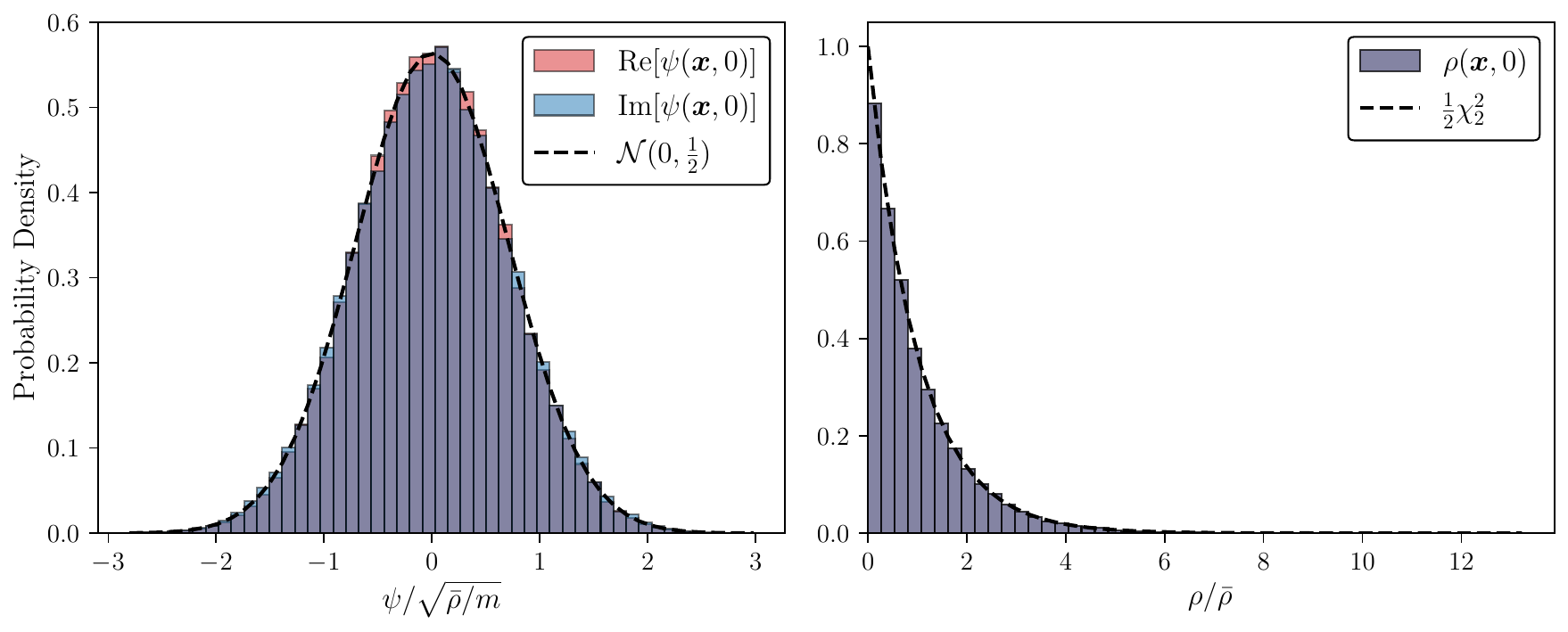}
    \caption{Histograms for the initial real/imaginary components of the dimensionless field $\psi({\bx,0})/\sqrt{\bar{\rho}/m}$ (left) as well as dimensionless density $\rho({\bx,0})/\bar{\rho}$ (right) for single realization of the field (e.g.\ Fig.~\ref{fig:initial-condition}). The field components match the desired  Gaussian normal distribution $\mathcal{N}(\mu,\sigma^2)$ with zero mean $\mu=0$ and variance $\sigma^2=1/2$ to good approximation. Consequently the density is  chi-squared distributed with $2$ degrees of freedom and unit normalization $\chi_2^2/2$.}
    \label{fig:statistics}
\end{figure*}

\subsection{Numerical Implementation}
\label{sec:numericalimplemetation}

In order to track the field evolution beyond perturbation theory and to test its domain of validity, we conduct fully-nonlinear numerical simulations.  
We employ the well-known pseudo-spectral algorithm \cite{Edwards:2018ccc,Jain:2022agt}, adapted to include the effects of the self-interaction potential $V_\SI$ (see \cite{Glennon:2020dxs,Painter:2024rnc,Mirasola:2024pmw}). The basic scheme is as follows: 1) evolve the wavefunction a half-step in time using the current non-linear potential, 2) evolve the wavefunction a full time step according to its free Hamiltonian, 3) evolve the wavefunction another half step according to the non-linear potential at this new time.
Schematically, we have:
\begin{equation}\label{eq:kick-drift}
    \begin{split}
    \psi({\boldsymbol{x}}, t+\Delta t) \approx & \exp \left[-\frac{i \Delta t}{2} V_{\rm SI}({\boldsymbol{x}}, t+\Delta t)\right] \\
    &\times  \mathcal{F}^{-1}\{\exp \left[{-i \Delta t} \frac{k^2}{2m}\right] \} \\ &\times \exp \left[-\frac{i \Delta t}{2} V_{\rm SI}({\boldsymbol{x}}, t)\right] {\psi}({\boldsymbol{x}}, t)
    \end{split}
\end{equation}
for some small timestep $\Delta {t}$, where  $\mathcal{F}^{-1}$ is the inverse Fourier transform and it is understood that all quantities are dimensionless in code units as discussed below.

We restrict our simulations to the $t_\dB \ll t_\vg \ll t_\therm$ case where $|\epsilon_g|\ll1$ through a suitable choice of $\kp$ and $g\brho/m$.    We choose the time range $t\leq 10t_{\therm}$ in order to capture the onset of fully nonlinear effects but avoid the strongly coupled regime where $t \gtrsim 10t_\therm$ and objects such as solitons or condensates form.    We again ignore gravitational interactions entirely which should be a good approximation as long as $t\ll t_\GN$.  

In \S \ref{sec:PT}, we showed that in perturbation theory, the scaling behavior of perturbations is the same in $n=3$ and $n=2$ spatial dimensions.
To reduce computational complexity and allow for iteration over a range of physical parameters, we choose $n=2$ here.   
Explicitly, we assume that the wavefunction still lives in all three spatial dimensions ${\vec x}= \{\bx,z\}$ but the initial conditions depends only on two $\psi_i({\vec x}) = \psi_i({\bx}$) for all $z$.   This symmetry is  preserved by the Schr\"odinger equation, which then can be solved in 2 spatial dimensions.

As such, we run the simulations over the 2-dimensional domain $[-\frac{L}{2},\frac{L}{2}]\times[-\frac{L}{2},\frac{L}{2}]$, with side length $L=1$ in code units, and $N=256$ grid points along each axis. For the  physical time step given this grid scale in physical units $\Delta x$, we require
\begin{equation}
    \frac{\Delta t}{m} \lesssim \frac{\Delta x^2}{2\pi},
\end{equation}
which can be interpreted as the free propagation timescale for the Nyquist mode across the grid scale. The field is scaled by $\sqrt{\bar{\rho}/m}$ to make its code units dimensionless and this value is fixed across all simulations.  The consequence is that our simulations are parameterized by $\epsilon_g$ and all lengths and timescales can be characterized in units defined by $\kp$, e.g. $t_\dB$ for time, so that the mass $m$ is arbitrary.

Finally, we take periodic boundary conditions such that
\begin{align}
    {\psi}(x+L,y,t)&={\psi}(x,y,t) \\
    \psi(x,y+L,t)&=\psi(x,y,t),
    \end{align}
as is typical with particle N-body simulations.
As with particle dark matter, this should be a good approximation for the dynamics of modes that are much smaller than the simulation volume assuming that those modes carry the dominant fluctuation and the simulations tile a larger domain that is statistically homogeneous.

Let us now discuss our choice of initial condition. Even if  $\psi$ only possesses  short wavelength contributions, $\rho$ possesses long wavelength contributions from field interference effects through Eq.~(\ref{eq:densitymode}), regardless of gravity and self-interactions.   In order to isolate these interference effects without an excessive dynamic range of spatial scales,
we choose an initial condition where the field has narrow support around a peak momentum mode $\ks$, which plays the role of the fiducial momentum scale of the previous sections  $\kp=\ks$.   Specifically we take 
\begin{equation}\label{eq:initial-condition}
   \psi_i(\bk) \propto 
e^{ - \frac{(k-\ks)^2}{2\sigshell^2} + i \alpha_{\boldsymbol{k}}},
\end{equation}
with the Gaussian shell width $\sigshell \ll \ks$.   The proportionality coefficient is chosen to reproduce the average density $\brho$.
The resulting density spectrum at low momentum determined from Eq.~(\ref{eq:densitymode}) can therefore  be attributed solely to the interference of pairs of high-$k$ modes $m (\psi_i^*(\bk_2)\psi_i(\bk_1)+ \cc)$
in a squeezed triangle configuration: $\bk = \bk_2-\bk_1$ where $k \ll k_1 \approx k_2 \approx \ks$ as there is no support in the initial wavefunction for these $k$-modes.

\begin{figure}
    \centering
    \includegraphics[width=0.99\linewidth]{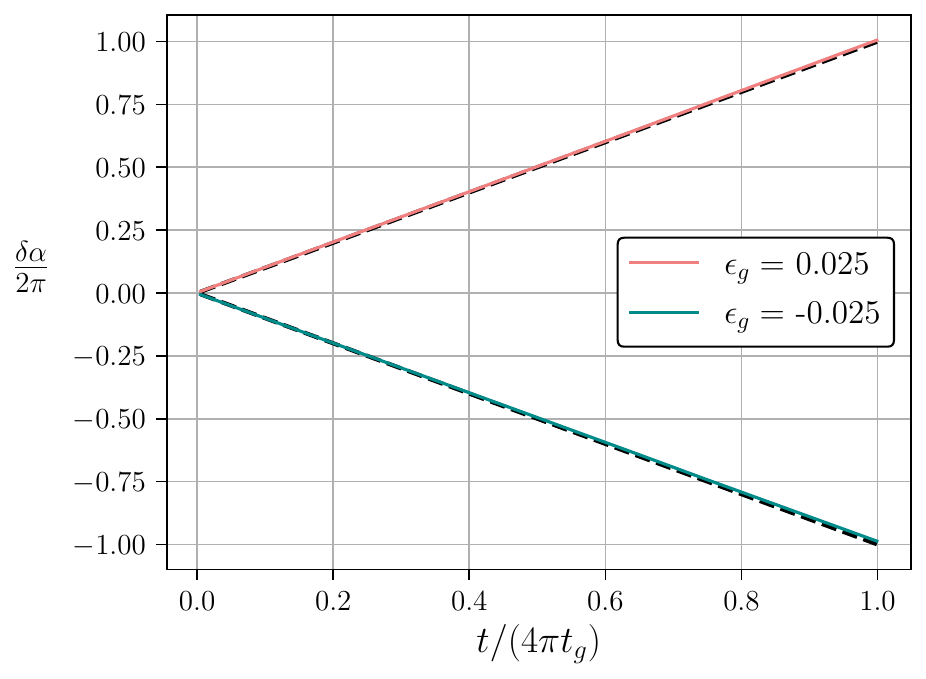}
    \caption{ Evolution of the self-interaction-induced phase contribution $\delta\alpha$ for $|\epsilon_g| = 0.025$, computed as an average over the well-occupied field modes in the bin $k/\ks\in[0.986, 1.006)$. Dashed black lines correspond to the theoretical prediction for the unperturbed evolution due to $\brho$ in Eq.~\eqref{eq:delta-alpha}.}
    \label{fig:phase-evolution}
\end{figure}

\begin{figure}
    \centering
    \includegraphics[width=0.99\linewidth]{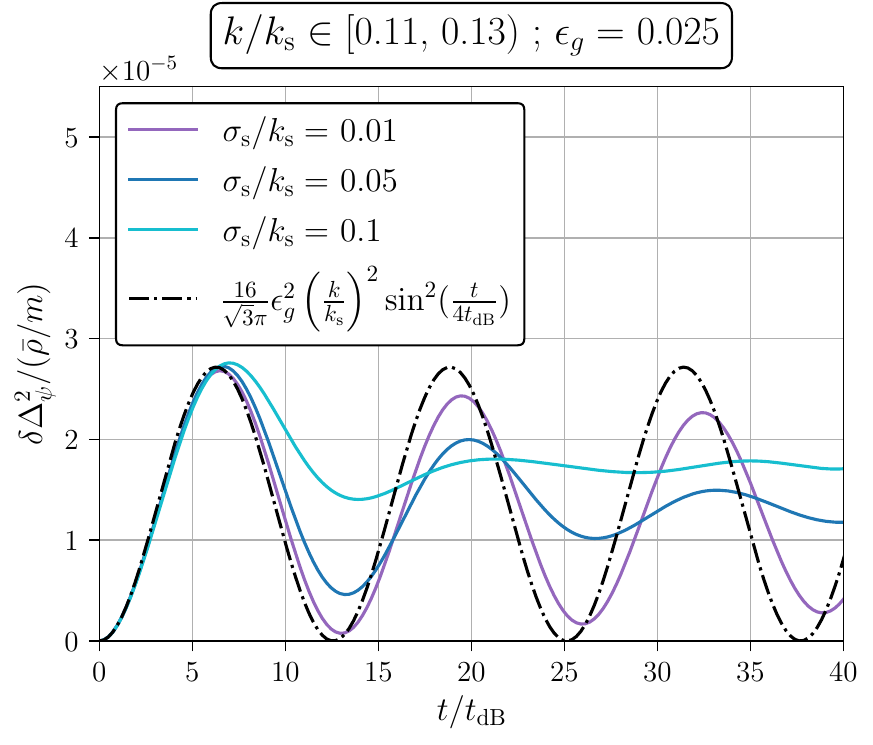}
    \caption{Comparison of generated low-$k$ field power from numerical simulations with the perturbation theory expression Eq.~\eqref{eq:field-power-theory}. A range of shell widths are simulated in order to demonstrate convergence towards the $\sigshell/\ks \rightarrow 0$ perturbation theory expression for $k/\ks=0.12$. Numerical power here is computed as an average over $k$ modes in the interval $k/\ks \in[0.11,0.13)$ and over $25$ simulations. Here the meaning of $\delta\Delta^2_{\psi}$ is that we are subtracting the $\Delta^2_{\psi}(\epsilon_g=0)$ value. }
    \label{fig:field-power-evolution}
\end{figure}

Specifically, after restoring the normalization factor $\brho/m$, the field modes are drawn from the initial power spectrum
\begin{equation}\label{eq:Pi-shell}
P_i(k) = \frac{ 4\pi\brho e^{(k-\ks)^2/\sigshell^2} }{
m \sigshell [\sqrt{\pi} \ks + e^{-\ks^2/\sigshell^2} \sigshell + \sqrt{\pi} \ks {\rm Erf}{(\ks/\sigshell)}]},
\end{equation}
with random phases 
 $\alpha_{\boldsymbol{k}}\in [-\pi,\pi)$  from a uniform distribution\footnote{Even if the phase were initially coherent, the free propagation of waves would rapidly produce an effectively random distribution on the $t_\dB$ timescale (e.g.~\cite{Liu:2024pjg}).}  
but fixed amplitude.
 We do not allow  stochasticity in the amplitude of $\psi_i(\bk)$ so as  eliminate cosmic variance on the initial field power spectrum and isolate interference effects. 
In the simulation context, ensemble average brackets $\langle...\rangle$ denote averages over many of these random phase simulations.
 
The cosmic variance of the initial density power spectrum in a single simulation volume remains, since the density spectrum is generated from the random phase interference of the field modes pairs, especially at low $k$.   
The exception is the $k=0$ mode where the field mode pairs have the same momentum and the spatial average and ensemble average density coincide under these assumptions, $\brho = \langle\rho\rangle$.
When comparing the simulations to the power spectra predictions, we average over many simulations to reduce cosmic variance and for various $\epsilon_g$ we difference simulations with the same initial field phases.

Note that in two dimensions
\begin{equation}\Delta_\psi^2 \equiv\frac{ k^2 P(k)}{2\pi}
\end{equation}
has the same dimensions as $|\psi(\bx,t)|^2$ and characterizes the field variance contributed per efold in $k$
\begin{equation}\label{eq:delta-rms}
    \psi_\rms^2(t)=\langle |\psi(\bx,t)|^2\rangle = \int \frac{dk}{k}  \Delta^2_{\psi}(k,t),
\end{equation}
and we will use this measure of the power spectrum in examples below.
Likewise 
\begin{equation}
\Delta_\delta^2(k,t) \equiv \frac{1}{\brho^2} \frac{k^2 P_\rho(k,t)}{2\pi}
\end{equation}
is the contribution to the real space  density variance per efold in $k$
\begin{equation}\label{eq:delta-rms}
    \delta_\rms^2(t)=\langle \delta^2(\bx,t)\rangle = \int \frac{dk}{k}  \Delta^2_{\delta}(k,t).
\end{equation}
We will use  $\delta_\rms(t)$ and $\Delta^2_\delta(k,t)$ to highlight the time and wavenumber dependence of the density field respectively.

An example initial field power spectrum and density field  is pictured in Fig.\ \ref{fig:initial-condition}.  Note that the peak of the power spectrum scales as
$\Delta_\psi^2 \sim (\ks/\sigshell) \brho/m$ and  $\delta_\rms \approx 1$ but overdensities are more common than underdensities reflecting the non-Gaussian statistics of the density field.  In Fig.~\ref{fig:statistics}, we show even though in momentum space we have only random phases and not random amplitudes, in real space we have a nearly Gaussian 1-point distribution function for the real and imaginary parts of $\psi$ and consequently the density field is $\chi^2$ distributed with 2 degrees of freedom with a long tail to $\rho/\bar\rho>1$.

\subsection{Time Evolution}

\label{sec:timedomain}

We now compare simulation results to the perturbation theory developed in Sec.~\ref{sec:PT} 
for the thin shell momentum distribution in two dimensions of Eq.~(\ref{eq:initial-condition}) and in the wave-interference dominated regime $|\epsilon_g|\ll1$.

Eq.~\eqref{eq:psi0} predicts an interaction sign-dependent but $k$ independent contribution to the field phase. In the perturbative regime, this gives the zeroth order effect for the initially occupied modes around $\ks$:
 \begin{equation}
 \frac{\psi(\bk,t;\epsilon_g)}{\psi(\bk,t;0)}=
 e^{-i\delta\alpha},
 \end{equation}
 where
 \begin{equation}\label{eq:delta-alpha}
 \delta\alpha = \frac{ g\brho}{2m} t= \frac{\mathrm{sgn}(\epsilon_g)}{2}\frac{t}{t_\vg}.
 \end{equation} 
In Fig.~\ref{fig:phase-evolution}, we verify this prediction with the simulations for modes around $\ks$ and $\epsilon_g=\pm 0.025$. We take a small value here so as to increase the separation  $t_\dB/t_\vg=|\epsilon_g|$ to isolate stages of the evolution in the power spectra below.
Notice that for $\delta\alpha$ we do not actually require $|\epsilon_g|\ll 1$ nor a small value for this phase shift, as long as field perturbations themselves remain small.  We have verified this prediction with a range of larger $\epsilon_g$ values.

The impact of the first order field perturbations can be seen in the field power $P(k)$ for modes with highly suppressed initial occupation, i.e.~$k\ll \ks$.  We can analytically integrate the $\mathcal{O}(g^2)$ correction $P_{11}(k)$ in Eq.~\eqref{eq:P11} using the approximation in 
Eq.~(\ref{eq:P11integral}) and the initial power spectrum~(\ref{eq:Pi-shell})  to obtain
\begin{equation}\label{eq:field-power-theory}
\delta \Delta_\psi^2=
    \frac{k^2 P_{11}}{2\pi} \approx \frac{\brho}{m} \frac{16}{\sqrt{3}\pi} \epsilon_g^2 \left(\frac{k}{\ks}\right)^2 \sin ^2\left(\frac{t }{ 4 t_{\dB}}\right).
\end{equation}
Here we have assumed that $\sigshell/\ks\ll 1$ and used the fact that 
\begin{equation}
\label{eq:deltafnshell}
    \lim_{\sigshell\rightarrow 0}P_i(k)=(2\pi)\delta(k-\ks)\frac{1}{k}\frac{\bar{\rho}}{m}.
\end{equation}

\begin{figure}
    \centering
    \includegraphics[width=0.99\linewidth]{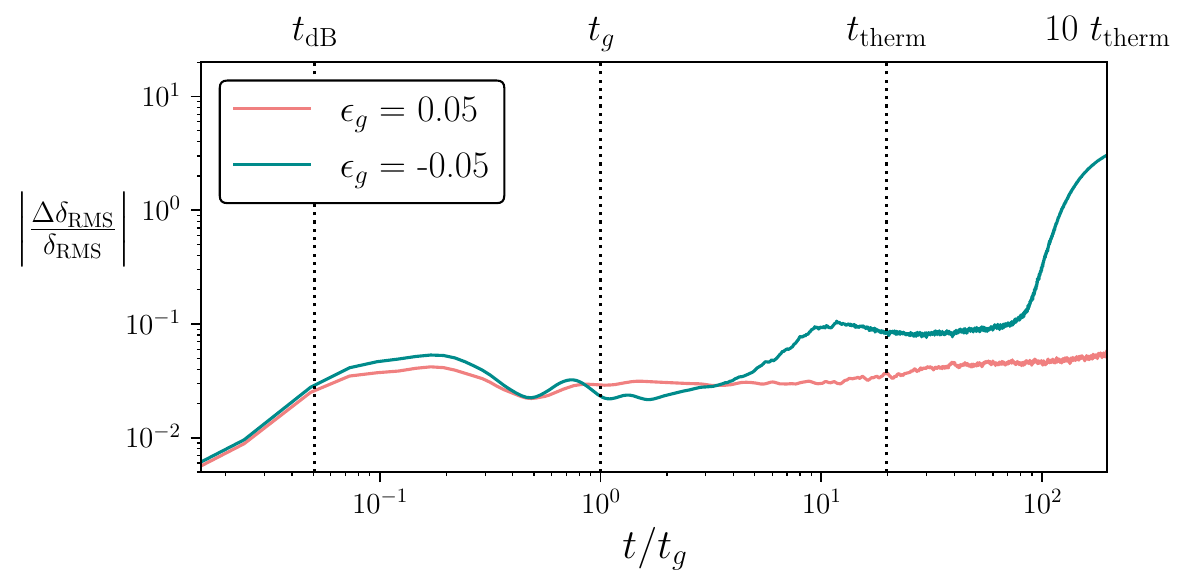}
    \caption{
    Fractional change $|\Delta\delta_\rms/\delta_\rms|$ with respect to the $\epsilon_g=0$ evolution for shell width $\sigshell/\ks=0.05$ and for times $t\leq 10 \ttherm$. $\delta_\rms^2$ here has been averaged over $50$ simulations.}
    \label{fig:delta-rms-evolution}
\end{figure}

\begin{figure*}
    \centering
    \includegraphics[width=0.99\linewidth]{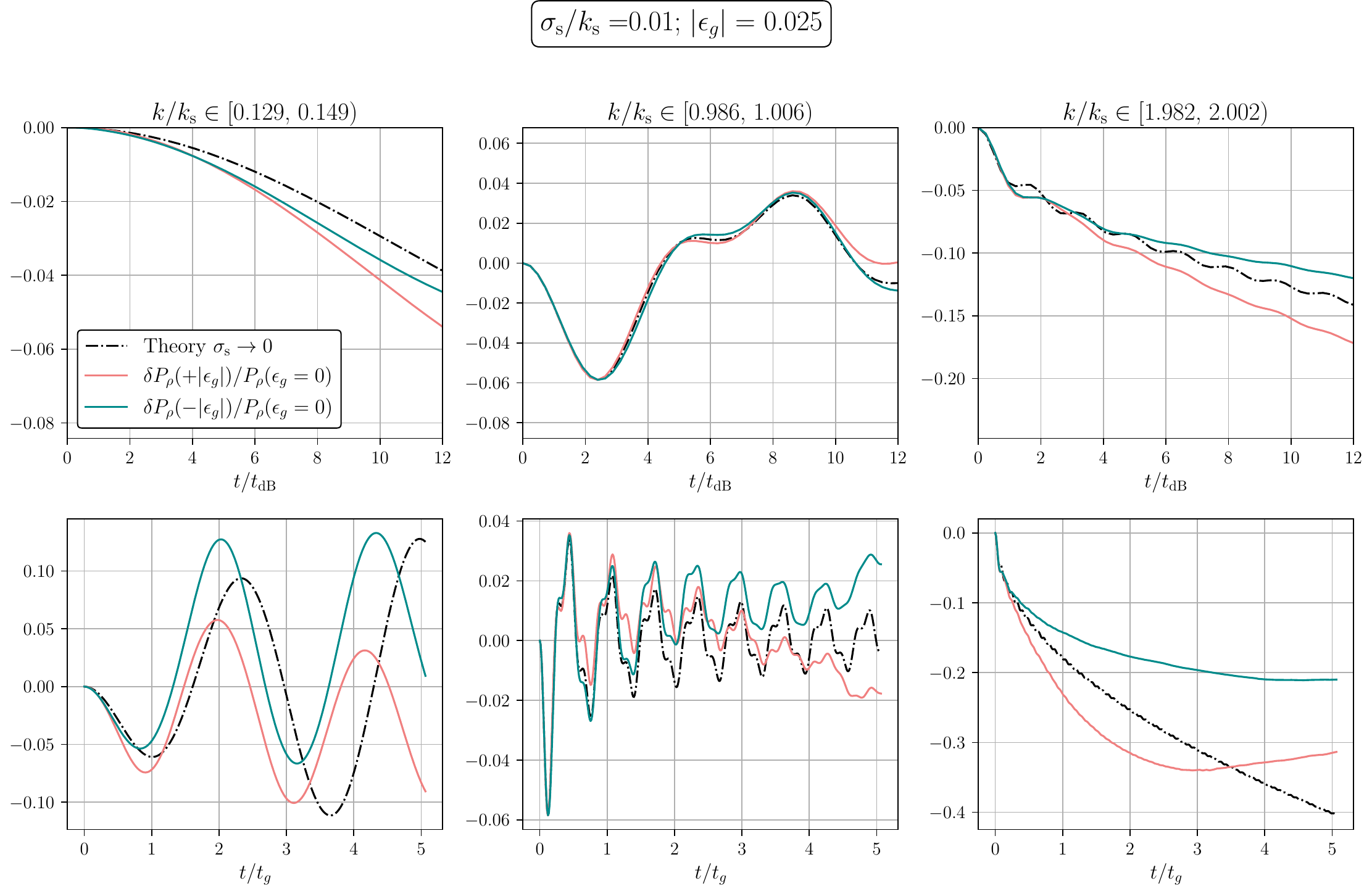}
    \caption{Fractional change in density power for $\epsilon_g=\pm 0.025$ compared to the perturbation theory result Eq.~\eqref{eq:deltaPrhodeltafn} (dashed curves).
    For easy comparison of amplitudes, we plot ${\rm sgn}(\epsilon_g)\delta P_{\delta}/P_{\delta}$. The top row shows the evolution at early times (in units of $\tdB$) whereas the bottom row shows the later-time dynamics (in units of $t_\vg$), where $\mathcal{O}(\epsilon_g^2)$ which distinguish attractive and repulsive interactions in amplitude becomes. For theory curves, we take $\sigshell\rightarrow 0$ and evaluate wavenumbers $k=0.12\ks,1\ks,2\ks$ respectively (see text for the interpretation of the $2\ks$ mode). Numerical results shown here are averaged over $25$ simulations. 
    } 
    \label{fig:density-power-evolution}
\end{figure*}

We compare these $\sigshell/\ks \rightarrow 0$ predictions against simulations in  Fig.~\ref{fig:field-power-evolution} for a range of $\sigshell$.
For all cases, the early time evolution $t\ll 4t_\dB$ matches the prediction well and reflects the expected quadratic growth.  For the smallest width tested $\sigshell/\ks=0.01$, the predicted saturation due to wave interference is also well matched by the simulations for the first few cycles. For larger widths up to $\sigshell/\ks=0.1$, saturation still occurs around $k\sim 4t_\dB$ but with more prominent deviations from a pure $\sin^2$ oscillation.   These occur due the interference phase $\Delta k^2$ varying due to the $\sim \pm\sigshell$ range of $k_{a,b,c}$ initial modes around $\ks$ and the $k$ mode in question (see Eq.~\ref{eq:Deltak2}).  Notably a decrease in $\Delta k^2$ causes a decrease in the temporal frequency of the oscillation as well as a prolonged period of growth.   The net effect after integrating over $k_{a,b,c}$ is a smoothing of the oscillations and a net decrease in the temporal frequency.  For the highest width $\sigshell/\ks=0.1$ there are even modes near $1\sigshell$ of $\ks$ where $\Delta k^2\rightarrow 0$ since 
$\Delta k^2 \rightarrow -\ks^2(1-6\sigshell/\ks)$ and contribute to a strong distortion of the evolution.  These can in principle be predicted by integrating the full expression~\eqref{eq:P11} numerically.
Even for $\sigshell/\ks ={\cal O}(1)$, 
the quadratic growth ceases after $t=4 t_\dB$ though the initially occupied modes within the shell continue to evolve linearly across the $t_\vg \lesssim t \lesssim t_\therm$ range as predicted by Eq.~(\ref{eq:thermtwiddle}).

\begin{figure*}[t]
    \centering
    \includegraphics[width=0.9\linewidth]{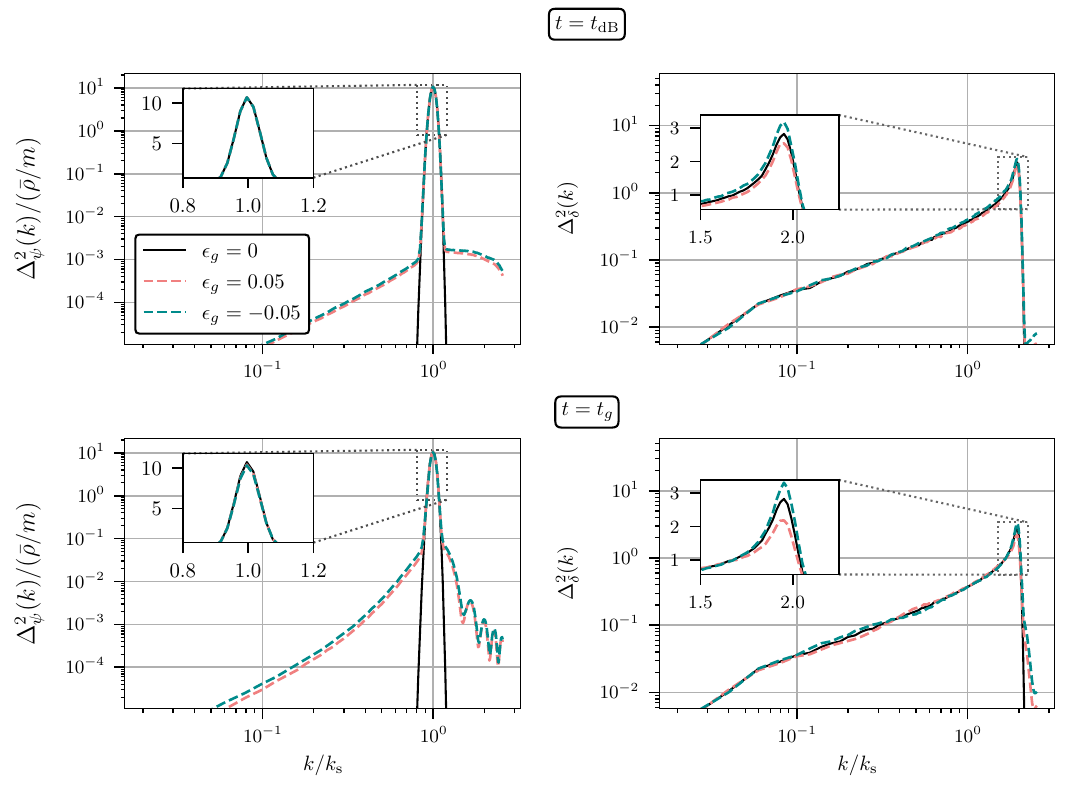}
    \caption{Comparison of dimensionless power spectra of field fluctuations (left column) and density fluctuations (right column) for a thin shell initial condition $\sigshell/\ks=0.05$ and three couplings: $\epsilon_g=0,\pm 0.05$. Power here is computed from snapshots taken at early times $t_{\rm dB}$ (top row) and $t_\vg$ (bottom row) and averaged over $50$ simulations.}
    \label{fig:power-early}
\end{figure*}

\begin{figure*}[t]
    \centering
    \includegraphics[width=0.9\linewidth]{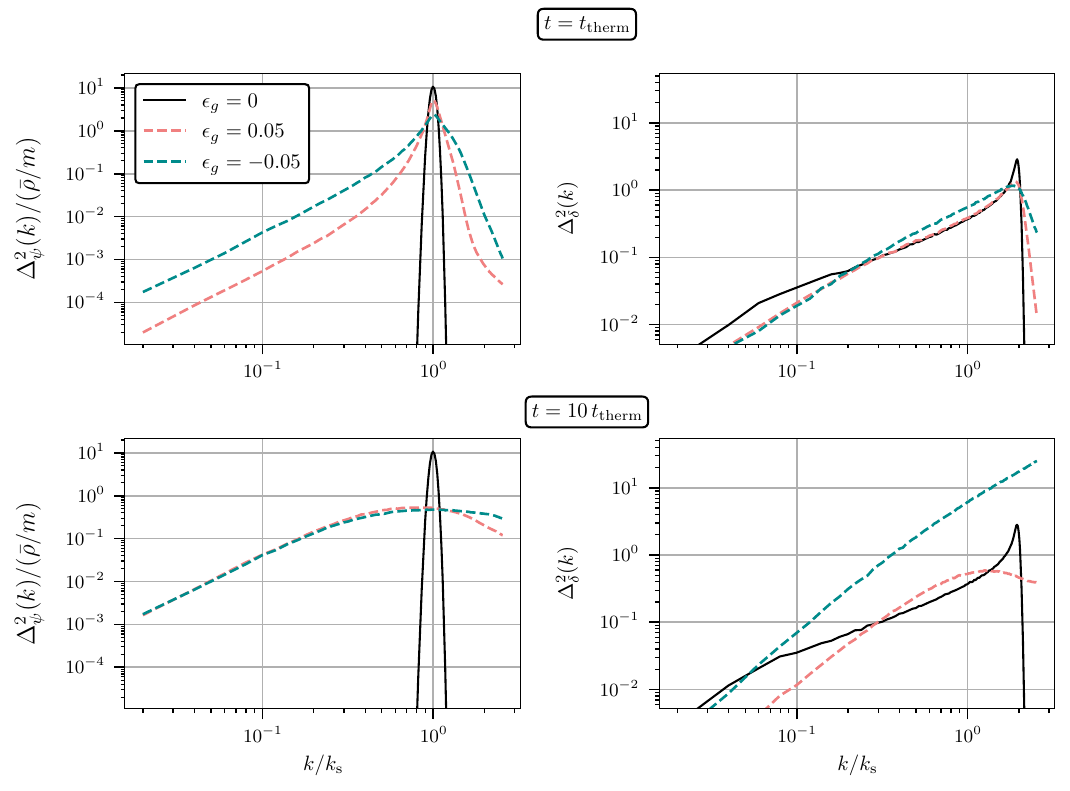}
    \caption{Comparison of dimensionless power spectra of field fluctuations (left column) and density fluctuations (right column) for a thin shell initial condition $\sigshell/\ks=0.05$ and three couplings: $\epsilon_g=0,\pm 0.05$. Power here is computed from snapshots taken at late times $\ttherm$ (top row), $10\ttherm$ (bottom row) and averaged over $50$ simulations. By $10\ttherm$, solitons have formed under the attractive interaction $\epsilon_g=-0.05$ whose true size is no longer resolved by the grid so that the power peaks at the Nyquist frequency. }
    \label{fig:power-late}
\end{figure*}

\begin{figure*}[t]
    \centering
    \includegraphics[width=0.9\linewidth]{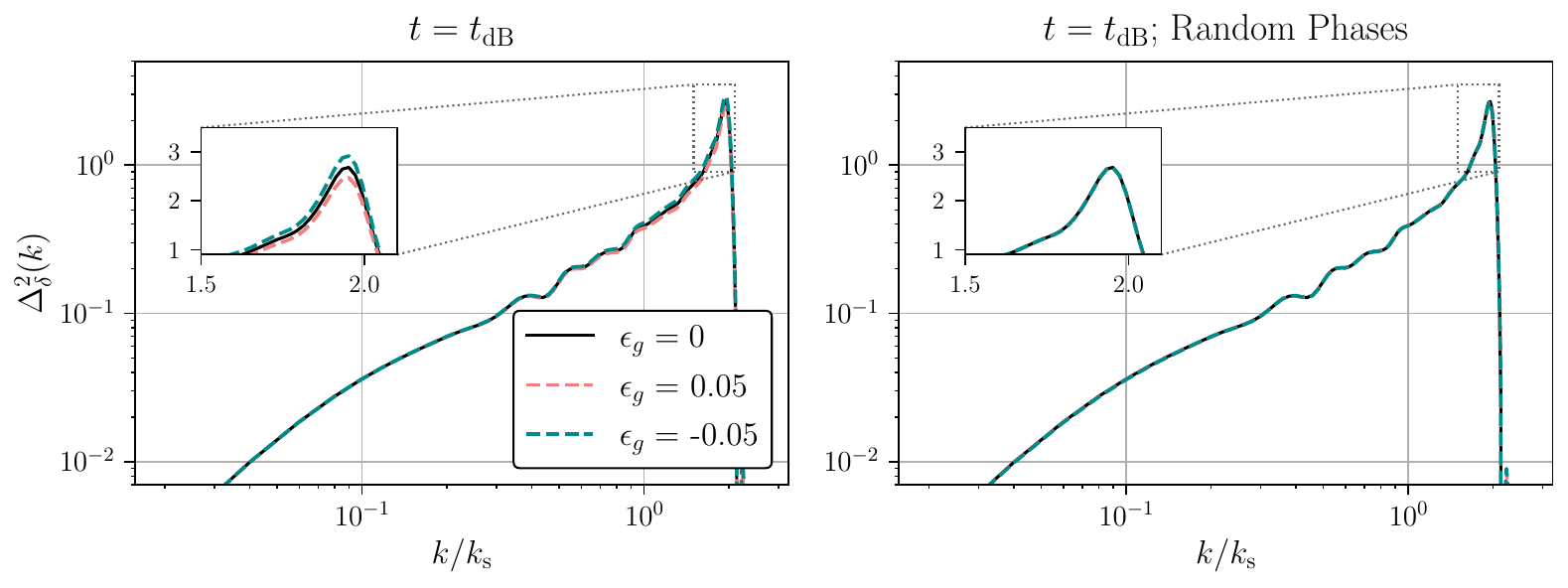}
    \caption{Density power spectrum computed at time $\tdB$ (left) versus a density power spectrum generated from field amplitudes $|\psi(\bk,\tdB)|$ but phases re-randomized.  The incoherence of the wavemodes destroys the change in power. The initial shell width is $\sigshell/\ks=0.05$.}
    \label{fig:random-phase-test}
\end{figure*}
%
%
%
\begin{figure*}[t]
    \centering
    \includegraphics[width=0.9\linewidth]{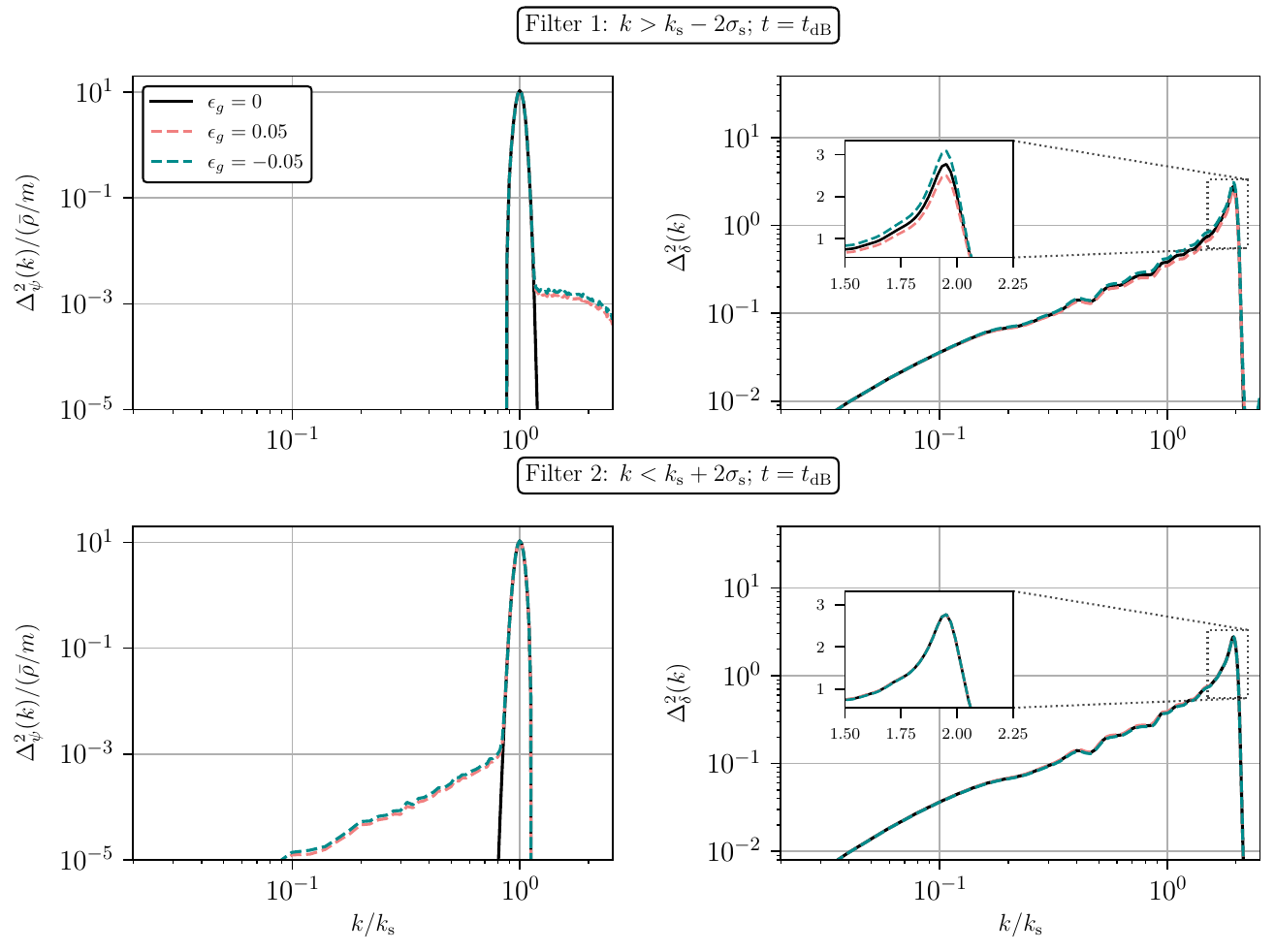}
    \caption{Resulting power spectra computed after applying high-pass (top row) and low-pass (bottom row) filters to the field $\psi(\bk,\tdB)$, both including the shell modes.  The perturbed field modes for the high-pass filter are mainly responsible for the change in the density power spectrum.
    }
    \label{fig:filter}
\end{figure*}

\begin{figure}
    \centering
    \includegraphics[width=0.99\linewidth]{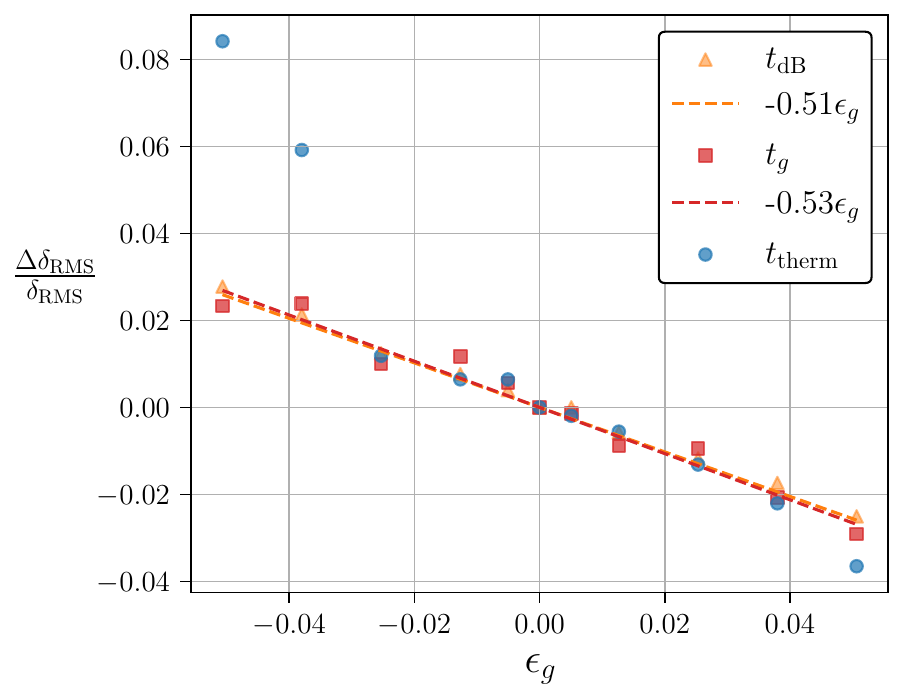}
    \caption{Fractional change in the RMS density contrast plotted as a function of the dimensionless coupling parameter $\epsilon_g$ for a thin shell of width $\sigshell/\ks=0.05$. We compare this scaling at times $t_{\rm dB}, \ t_\vg$ and $t_{\therm}$, and perform a linear fit for times $t_{\rm dB}$ and $t_\vg$ (dashed lines). Positive values of $\epsilon_g$ correspond to repulsive interactions, whereas negative values are attractive.}
    \label{fig:delta-RMS-scaling}
\end{figure}

We make similar comparisons  for the density power spectrum.   Again we can explicitly integrate the predictions for $k \ll \sigshell \ll \ks$ and the early time quadratic growth behavior in Eq.~(\ref{eq:deltaPrho}) as in Eq.~(\ref{eq:deltaPrholowk}).  Moreover, we can also integrate it for $k \gg \sigshell$ using the fact that this implies $|\bk + \bk_a|\approx \ks$ and $k_a\approx \ks$ or $2\bk_a \cdot \bk \approx -k^2$ for a thin shell.  In both limits, the initial growth follows
\begin{equation}
\lim_{t \ll m/k\ks} \delta P_\rho(\bk,t)
=
- \left( \frac{k}{\ks}\frac{t}{t_\SI} \right)^2  P_{\rho}(k,0)
\label{eq:deltaPrhoshell}
\end{equation}
and a numerical integration shows that this is in fact a good approximation across the full range of $k$ including $k \sim \sigshell$.  The phase incoherence of the modes makes the growth saturate when $t \gtrsim t_{\dB}(\ks/k)$ and so at a different timescale for different $k$ but at a comparable amplitude $(t_\dB/t_\SI)^2=\epsilon_g$,
\begin{equation}
\label{eq:deltaPrhosat}
\lim_{t \gg m/k\ks} \delta P_\rho(\bk,t)=\mathcal{O}(\epsilon_g) P_{\rho}(k,0).
\end{equation}
As a benchmark for comparison, we can again consider the delta function shell limit of Eq.~(\ref{eq:deltafnshell}) and obtain the initial density power spectrum 
\begin{equation}
\lim_{\sigshell \rightarrow 0} 
P_\rho(k,0) =
\begin{cases} \frac{4}{k\sqrt{4 \ks^2 -k^2}} \brho^2 & k<2\ks \\
0 & k\ge 2\ks
\end{cases}.
\end{equation}
Here $\delta_\rms(t=0)=1$ reflecting the ${\cal O}(1)$ density fluctuations expected from the initial field modes.

In this case we can predict the behavior from growth through saturation as a numerical integral
\begin{equation}
\label{eq:deltaPrhodeltafn}
    \lim_{\sigshell \rightarrow 0}\frac{ \delta P_\rho(k, t)}{ P_\rho(k, 0) }
    = -\frac{16}{t_\vg^2}   \int \frac{d \theta}{2 \pi}
    \frac{m^2}{k_\theta^2} {\sin ^2\left(\frac{k_\theta^2 t}{4 m}\right)},
\end{equation}
where $k_\theta^2 = k^2+2 k \ks \cos \theta$.
Note that a similar expression holds for a $n=3$ dimensional shell and in that case the integral can be expressed in closed form in terms of the cosine integral.
These perturbation theory expectations should hold for $t\ll t_\therm = t_\vg/\epsilon_g$.

We compare these predictions to simulations in Fig.~\ref{fig:density-power-evolution}.  For  $t\lesssim \tdB$ and the full range of $k$, the simulations match the predicted quadratic growth closely and confirm that the saturation time decreases with increasing $k$.   Notice that at $k= 2\ks$ the perturbation theory prediction encounters a pole where $\Delta q^2$ and the phase coherence parameter in Eq.~(\ref{eq:deltaPrhodeltafn})  $k_\theta^2 \rightarrow 0$ for $\theta =\pi$.  This causes the growth to continue indefinitely.  In the $n=3$ case where there is a closed form expression, growth is logarithmic in time.

A stronger saturation occurs in the finite $\sigshell$ simulations, and we shall see, the $\sigshell$ dependence mainly reflects the redistribution of power from an initial spectrum with a sharp peak at $k=2\ks$ whereas $\delta_{\rms}$ reflects changes across the whole distribution.

Finally, notice that on timescales $t\gtrsim t_\vg$, $P_\delta$ begins to exhibit $\mathcal{O}(\epsilon_g^2)$ corrections, seen in the bottom row of Fig.~\ref{fig:density-power-evolution}, where $\pm |\epsilon_g|$ no longer predict the same amplitude for $|\delta P_\rho/P_\rho|$.  This is expected from the breakdown of perturbation theory approaching the thermalization time, since we expect 
\begin{equation}
\delta P_\rho(k,t) \sim \frac{t}{t_\therm} P_\rho(k,0)
\sim \frac{t }{t_\dB} \epsilon_g^2 P_\rho(k,0),
\end{equation}
which becomes comparable to the saturated linear order term of Eq.~(\ref{eq:deltaPrhosat})   around $t\sim t_\vg$.

We can see the full progression to thermalization in the RMS density fluctuation.   In this case we choose a larger $|\epsilon_g|=0.05$ to reduce the separation between scales to visualize the whole process.
At $t\ll t_\dB$, we see the quadratic growth expected from coherent phase evolution which then settles into a quasistatic phase with fractional amplitude ${\cal O}(\epsilon_g)$ as expected from the  saturation phase of perturbation theory through $t\sim t_\vg$.  Note that for comparison with the perturbation theory predictions, $\delta_\rms$ is dominated by the largest momentum  modes, $\ks \lesssim k \lesssim 2 \ks$, where the saturation time is $\sim t_\vg$.  

Between $t_\vg \lesssim t \lesssim t_\therm$, $\delta_\rms$ does not grow substantially but as discussed above and detailed in the next section, the same total power  gets redistributed in momentum, indicating the onset of thermalization.   At the final epoch $t= 10 t_\therm$ we see a large change in the attractive case $\epsilon_g < 0$ indicating the onset of soliton formation which we discuss in the next section.

\subsection{Field and Density Power Spectra} 
\label{sec:results}

To understand the momentum space distribution changes due to interactions, it is also instructive to plot the field and density power spectra $\Delta^2_\psi(k,t)$ and $\Delta^2_\delta(k,t)$ as a function of $k$ at  $t$ fixed to the characteristic times  discussed in the previous section.  In Fig.~\ref{fig:power-early}, we take these to be
$t_\dB$ and $t_\vg$, whereas in Fig.~\ref{fig:power-late} we take $t_\therm$ and $10t_\therm$. 
This delineation is chosen to represent the initial growth to quasistatic saturation phase and thermalization phase separately.

\begin{figure*}
    \centering
    \includegraphics[width=0.85\linewidth]{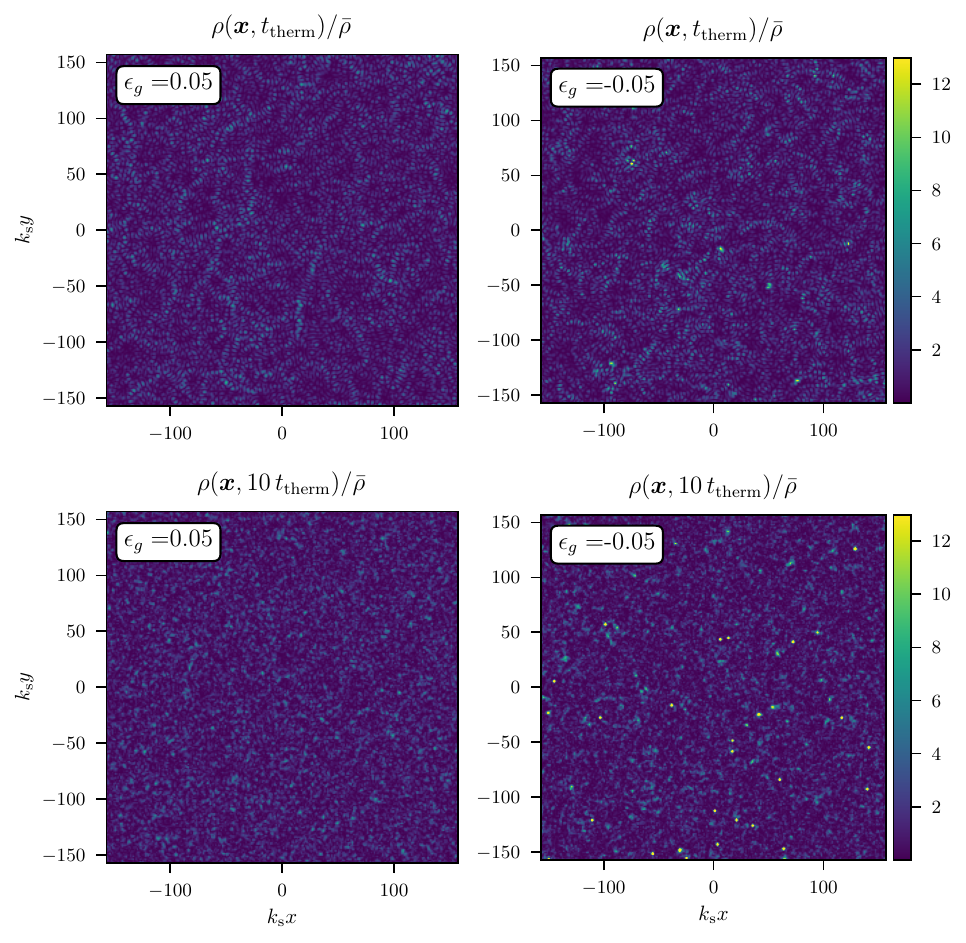}
    \caption{Snapshots of the real space density at times $\ttherm$ (top row) and $10\ttherm$ (bottom row), evolved with repulsive $\epsilon_g=0.5$ (left column) and attractive $\epsilon_g=-0.5$ (right column). For ease of comparison, we have clipped the maximum value of $\rho$ displayed to match that of Fig.~\ref{fig:initial-condition}, which highlights the onset of soliton formation in the bottom right panel. The initial shell width here is $\sigshell/\ks=0.05$.}
    \label{fig:snapshots}
\end{figure*}

By $\tdB$, there is an $\mathcal{O}(\epsilon_g)$ suppression/enhancement of density power near the peak mode $k\sim 2\ks$, persisting through $t_\vg$. However, there is no such accompanying effect in peak mode $\ks$ of the field power. Rather, for these early times, field evolution is characterized by $\mathcal{O}(\epsilon_g^2)$ generation of power away from the initial shell as predicted from perturbation theory through $P_{11}(k,t)$. 

In perturbation theory, this difference in behavior arises from the phase coherence between a single perturbed $\psi_1$ mode with three unperturbed $\psi_0$ modes for
$\delta \Delta^2_\delta$, whereas evolution in $\delta \Delta^2_\psi$ is only at higher order from the coherence of two perturbed $\psi_1$ modes (see Fig.~\ref{fig:modes}).  To test that $\delta \Delta^2_\delta$ arises from phase coherence effects, we take the amplitudes of the time-evolved field modes at $\tdB$ in the simulations, but re-randomize their phases $\alpha_{\bk}$. This preserves the growth of  $|\psi(\bk)|$, but destroys any possible  phase information that can coherently superimpose to form changes to the density. We then compare the density power as calculated from the field with and without the inclusion of the phase randomization, shown in Fig.~\ref{fig:random-phase-test}.  The result is that by removing the self-interaction induced phase, the $\mathcal{O}(\epsilon_g)$ effects on $\Delta^2_{\delta}$ are also cancelled out.

In order to  determine which modes contribute most to this coherent effect, we can instead filter out different modes of $\psi(\bk,\tdB)$. We apply both a high-pass filter $k>\ks-2\sigshell$ and low-pass filter $k<\ks+2\sigshell$, chosen such that the modes within the shell which contribute to $\rho_0$ are still included in either case. We then compute the field and density power, shown in Fig.~\ref{fig:filter}. From this, it is evident that the modes populated with   $\psi_1(k \gtrsim \ks)$ provide the dominant contribution, consistent with the perturbation theory expectation (see the discussion preceding Eq.~\eqref{eq:Prho-peak-scaling}).

Next, we return to the RMS density contrast to tests its  scaling with $\epsilon_g$. The time-evolution of $\delta_\rms$ is discussed in Sec.~\ref{sec:timedomain}. In Fig.~\ref{fig:delta-RMS-scaling}, we show that well before $t_\therm$ but after $t_\dB$, the scaling is linear as expected and does not appreciably evolve, indicating that wave interference prevents what would otherwise be a much larger evolution due to self-interaction.


Finally turning to the thermalization time scale, by  $\ttherm$, we see in the first row of Fig.~\ref{fig:power-late}, there is an $\mathcal{O}(1)$ suppression of the peak field power for \textit{both} attractive and repulsive interactions, reflecting the redistribution of power from the sharp initial peak at $k \approx 2\ks$.
For the attractive case, there is more power and by the end of the simulations at $10t_\therm$.   The power peaks at the highest $k$ and smallest scale resolved by the simulation with a nearly white noise tail, indicating the incipient formation of discrete bound objects or solitons.  In the snapshots from Fig.~\ref{fig:snapshots} we can see the correspondence in real space.   Here there are now discrete high density regions that are no longer transient interference phenomena but only in the attractive case $\epsilon_g<0$. This differs dramatically from the repulsive case, where in the limit $t\ll t_\GN$, there is no mechanism for the formation of solitons.

\section{Discussion}
\label{sec:discussion}

We have introduced wave perturbative techniques for \SIFDM\ and compared the predictions extensively against simulations in order to study the interplay between wave interference of modes and self-interaction.   For an initial field power spectrum characterized by some typical momentum scale $\kp$, we find that when the deBroglie timescale $t_\dB=m/\kp^2$ is shorter than the self-interaction timescale $t_\vg= m/g\brho$,  the relevant phases decohere and wave interference prevents both the redistribution of initial wave momenta and the change of their density fluctuations.  The ratio of timescales $t_\dB/t_\vg= |\epsilon_g|$ characterizes the quasistatic evolution in the decoherence regime.  In the absence of wave interference, density fluctuations would double on the timescale $t_\SI (\kp/k)= m/k \sqrt{g\brho}$  but instead with interference, they saturate at an $\epsilon_g$ suppressed value for $|\epsilon_g|<1$. 
 If  on the other hand $|\epsilon_g| >1$,  then self-interaction growth will occur before wave decoherence can act to stop it.  We characterize the perturbative regime with simple integrals over the initial field power spectrum, give the separate decoherence criteria for the field and density evolution, and show that they agree with simulations in their domain of validity.

These results have wide ranging application to \SIFDM\ phenomena which span a large range in time and physical scales.  For the QCD axion dark matter, due to the temperature dependence of the mass, misalignment and string network field fluctuations after inflation begin in the low momentum $|\epsilon_g|>1$ limit,  where self interactions change both the momentum distribution and density fluctuations of the dark matter.  As the Universe expands they then evolve into the $|\epsilon_g|<1$ limit where wave phenomena dominate on small scales and this evolutionary sequence can greatly alter the predictions for small scale structure \cite{Gorghetto:2024vnp}.  Other dark matter models where production occurs in causal domains after inflation may have their fluctuations originate directly in the ``warm" or high momentum regime where $|\epsilon_g|<1$ \cite{Amin:2022pzv,Liu:2024pjg}.

For \SIFDM\ scenarios where the interaction plays a direct role in the late time evolution of dark matter substructure, $\epsilon_g$ can also range widely over physical scales.   In these models, a repulsive interaction can act to stabilize dark matter against gravitational collapse on small scales.  In the absence of wave interference, if we set the self-interaction time scale for density growth equal to  the dynamical time
\begin{equation}
\frac{m}{k \sqrt{g\brho}} = \frac{1}{\sqrt{G\rho}}
\end{equation}
this picks out a physical scale $k^{-1}$ where repulsive interactions form cores to dark matter halos.  Restoring order unity factors, the core scale is given by
\cite{Khlopov:1985jw,Goodman:2000tg,Peebles:2000yy,Dawoodbhoy:2021beb}
\begin{eqnarray}
R_{\rm TF} &=& \pi \sqrt{\frac{g}{4\pi G m^2}}\nonumber\\
&=&0.79 \sqrt{ \frac{g/m^2}{10^{-18} {\rm cm}^3/{\rm eV}}} {\rm kpc},
\end{eqnarray} 
which is notably independent of the density and sometimes called the Thomas-Fermi scale.
However this consideration applies in the absence of wave interference phenomena.  To assess its validity, we can roughly estimate $\epsilon_g$ in dark matter substructures.  Following Ref \cite{Wolf:2009tu}, given a measured stellar velocity dispersion $\sigma_*$ of dwarf satellite halos, we can estimate the total mass as 
\begin{equation}
M = 3 \sigma_*^2 r_{1/2}/G
\end{equation}
where $r_{1/2}$ is the half-light radius.  Taking halo extent to be of order the half light radius itself, we obtain an estimate of the dark matter density 
\begin{equation}
\rho_{1/2} = \frac{M}{4\pi r_{1/2}^3/3} = \frac{9}{4} \frac{\sigma_\star^2}{\pi r_{1/2}^2 G}.
\end{equation}
Then 
\begin{equation}
\epsilon_g = \frac{g \rho_{1/2}}{m^2 \sigma_{\rm dm}^2} =  \frac{9}{\pi^2} \left( \frac{\sigma_\star}{\sigma_{\rm dm}}\frac{R_{\rm TF}}{r_{1/2}} \right)^2 
\end{equation}
where we have re-expressed the characteristic momentum with the dark matter velocity dispersion $\kp = m\sigma_{dm}$.
For example for ultra faint dwarf (UFD) galaxies that have been used to constrain the FDM mass with stellar heating
\cite{Dalal:2022rmp}, $r_{1/2}$ = 50 pc,
$\sigma_*$ = 3 km/s, 
and $\sigma_{\rm dm}$ = 6 km/s whereas a very different scale for $R_{\rm TF}\sim 1$kpc has been proposed to ameliorate the too-big-to-fail issue \cite{Dawoodbhoy:2021beb}, though it may be resolved within the cold dark matter paradigm itself.  More generally given a choice of $R_{\rm TF}$ which sets the self-interaction core, halos with larger half-light radius should be in the $|\epsilon_g|<1$ regime and halos with smaller, if they form at all and are consistent with the observed density profile, in the $|\epsilon_g|>1$ regime.  For the former case, UFD stellar heating bounds on the FDM mass should apply equally well to \SIFDM\ whereas for the latter heating could be substantially altered by rapid self-interaction.    In either case, sufficiently away from the center of the halo, the density  drops as does $|\epsilon_g|$.  Once $|\epsilon_g|<1$ in the outskirts, \SIFDM\ constraints from stellar dynamics become equivalent to FDM in this respect
\cite{Bar-Or:2018pxz,Amin:2022pzv,Yang:2024hvb, Hamilton:2024xeg}.
We leave the question of how constraints are modified in the $|\epsilon_g|>1$ regime to a future study.

Similarly our techniques apply to the interplay of wave interference and gravitational interactions which provides the wave analogue of free streaming suppression of the growth of structure.  Our wave perturbative techniques therefore can be used to reveal a rich range of phenomena where interference plays a role.

\vfill

\acknowledgements
We thank Andrey Kravtsov, Rayne Liu and Huangyu Xiao for useful conversations.
 C.C. is supported in part by the Arthur B. McDonald Institute via the Canada First Research Excellence Fund and by a Doctoral Research Scholarship
from the Fonds de Recherche du Qu\'ebec – Nature et Technologies.
W.H. is supported by U.S. Dept.\ of Energy contract DE-FG02-13ER41958 and the Simons Foundation.
E.M. is supported in part by a Discovery Grant from the Natural Sciences and Engineering Research Council of Canada, and by a New Investigator Operating Grant from Research Manitoba.


\bibliographystyle{apsrev4-1}
\bibliography{references}


\end{document}